\documentclass{article}
\usepackage{amssymb,amsfonts,amsmath,latexsym}
\def\operatorname#1{\mathop{\rm#1}\nolimits}
\def\tr{\operatorname{tr}}
\def\const{\mbox{const}}
\def\sgn{\operatorname{sgn}}
\def\tanh{\operatorname{tanh}}
\def\cosh{\operatorname{cosh}}
\def\sinh{\operatorname{sinh}}
\def\rk{\operatorname{rk}}
\def\Im{\operatorname{Im}}
\def\Re{\operatorname{Re}}
\def\ROM#1{{\uppercase\expandafter{\romannumeral#1}}}
\def\bad{\spaceskip=0.33emplus0.6emminus0.15em\immediate
        \write5{\string\bad}}
\def\tag #1 {\eqno{(#1)}}
\def\procl #1~#2{\begin{#1}\def\endproclaim{\end{#1}}}

\def\proclaim#1{\procl#1}

\def\demo#1{\addvspace\medskipamount\noindent{\sc#1}}
\def\enddemo{\hfill$\Box$\newline\smallskip}
\def\text{\mbox}
\def\script #1{{\cal #1}}
\def\definition#1{\par\medskip\noindent{\bf#1}}
\def\remark#1{\par\medskip\noindent{\bf#1}}
\def\enddefinition{\medskip\par}
\def\endremark{\medskip\par}
\def\rom#1{\noindent\/{\rm#1}}
\def\gather{\begin{aligned}}
\def\endgather{\end{aligned}}
\def\example#1{\addvspace\medskipamount\noindent{\bf#1}}
\def\endexample{\medskip}
\def\head\boldsymbol\S#1. #2\endhead{\section*{\S#1. #2}%
\addcontentsline{toc}{section}{\S#1. #2}}
\def\roster{\begin{itemize}}
\def\endroster{\end{itemize}}
\def\Item"#1"{\item[#1]}
\title{Discrete spectral symmetries of low-dimensional
differential operators and difference operators on regular
lattices and two-dimensional manifolds}
\author{S. P. Novikov and I. A. Dynnikov
\thanks{The work of the second author was
partially supported by the Russian Foundation
for Fundamental Research (grant no.~96-01-01404).
The last three sections of the survey present
the joint work of the authors announced in~\cite{1}.}}
\date{}

\begin{document}
\maketitle

\tableofcontents

\head
\boldsymbol\S1.
Introduction: history of the problem,
the one-dimensional Schr\"odinger operator
\endhead

Already in the 18th century (1742) Euler paid attention 
to the substitutions which transform solutions of a 
certain ordinary linear differential equation into 
solutions of another equation associated with it. Let
$$
L=-\partial_x^2+u(x)
\tag 1
$$
be the Sturm--Liouville operator defined on the real 
line~$(x)$, and let $\psi(x,\lambda)$ be a solution of 
the linear equation $L\psi=\lambda\psi$. Let $v(x)$ be 
an arbitrary solution of the\linebreak

Riccati equation
$$
u(x)=v_x+v^2.
\tag 2
$$
Under these conditions we have the following result.

\proclaim{Lemma~1}
The new function $\widetilde\psi=\psi_x-v(x)\psi$ 
satisfies the new equation $\widetilde L\widetilde\psi
=\lambda\widetilde\psi$, where
$$
\widetilde L=-\partial_x^2+\widetilde u(x),\qquad
\widetilde u(x)=-v_x+v^2.
$$
\endproclaim

\demo{Proof}
This lemma easily follows from the fact that $L$ is in 
fact the product of two non-commuting factors of the 
first order:
$$
L=-\partial_x^2+u=-(\partial_x+v)(\partial_x-v),
$$
where $u=v_x+v^2$. Further, we have
$$
-\psi_{xx}+u\psi=\lambda\psi=-(\partial_x+v)
(\partial_x-v)\psi=
-(\partial_x+v)\widetilde\psi.
$$
Hence,
$$
\widetilde
L\widetilde\psi=-(\partial_x-v)(\partial_x+v)\widetilde\psi=
\lambda(\partial_x-v)\psi=\lambda\widetilde\psi,
$$
where
$$
\widetilde L=-(\partial_x-v)(\partial_x+v)=
-\partial_x^2+\widetilde u(x).
$$
We have proved Lemma~1.
\enddemo

Hence, the Euler substitution is based on the 
representation of $L$ as the product of first-order 
operators which are formally adjoint to each other 
if $u,v\in\mathbb R$:
$$
L=-(\partial_x+v)(\partial_x-v)=QQ^+,
\tag 3
$$
and is given by interchanging them:
$$
QQ^+=L\longmapsto\widetilde L=Q^+\,Q,\qquad
\psi\longmapsto\widetilde\psi=Q^+\psi.
\tag 4
$$

In the literature the Euler substitutions are often 
called `Darboux transformations' or `B\"acklund--Darboux 
transformations' because of the role they play in 
the theory of non-linear systems of 
Korteweg--de~Vries~(KdV) type for functions 
$u(x,t)$ that depend on the time as the parameter.

The classical mathematicians put it like this: 
the Darboux transformation $L\mapsto\widetilde L$ 
is determined by one chosen solution $\varphi$ of 
the equation $L\varphi=0$, and
$$
\widetilde\psi=\psi_x-(\varphi_x/\varphi)\psi.
\tag 5
$$
We have
$$
v(x)=\varphi_x/\varphi, \quad\text{where}\quad 
v_x+v^2=u(x).
\tag 6
$$
The algebraic language, that is, the correspondence 
of this with the factorization $\alpha+L=QQ^+$, was 
used after the works of physicists of the 30s and 
40s (Dirac, Schr\"odinger, Infeld).

>From the point of view of formal spectral theory, 
that is, of local solutions of the equation $L\psi=
\lambda\psi$, the Darboux--Euler transformation 
generates a transformation of the eigensubspaces 
for all~$\lambda$:
$$
B_\lambda\:\psi\longmapsto\widetilde\psi,
$$
where $L\psi=\lambda\psi$, $\widetilde L\widetilde\psi=
\lambda\widetilde\psi$, $\widetilde\psi=\psi_x-v\psi$.

We should talk about the Darboux--Euler transformation 
of the whole linear hull (with respect to all~$\lambda$)
$$
B\:\sum a_i\psi_i\longmapsto\sum a_i\widetilde\psi_i,
$$
where $L\psi_i=\lambda_i\psi_i$ and $a_i$ are arbitrary 
coefficients.

The transformation $B$ is not, strictly speaking, an 
isomorphism: for $\lambda=0$ and $\psi=\varphi$ we have
 $B\varphi=0$, $v=\varphi_x/\varphi$. Hence, the kernel
  of this transformation is one-dimensional.

Turning to the present `global' spectral theory of the
operator $L$ in the Hilbert space $\script L_2(\mathbb
R)$, we see that if the function $\varphi(x)$ does not 
belong to $\script L_2(\mathbb R)$, then the Darboux 
transformation $B$ can generate an isomorphism of the 
spectral theories of the operators $L$ and~$\widetilde 
L$; sometimes $B$ is a monomorphism on a part of the 
spectrum of $L$ that covers all but one eigenfunction 
in $\script L_2(\mathbb R)$; sometimes (if 
$\varphi(x)\ne0$ and $\varphi\in\script L_2(\mathbb R)$) 
$B$~annihilates exactly one state of the spectrum 
$\varphi$, the lowest one, the `basic' one 
in~$\script L_2(\mathbb R)$. We do not know the 
complete classification of all possible cases from 
the point of view of spectral theory in~$\script 
L_2(\mathbb R)$.

\definition{Definition~1}
We call the Darboux--Euler transformation the {\it 
latent algebraic symmetry} of the spectral theory of 
the one-dimensional Schr\"odinger operator, or the 
{\it discrete symmetry}.
\enddefinition

We recall that the `discontinuous spectral symmetries' 
or `isospectral deformations' of the operator $L$ are 
the well-known systems of the theory of solitons 
like~KdV, which generate a gigantic commutative 
continuous group and in which KdV is a one-parameter 
subgroup, see~\cite{2},~\cite{3}.

The Darboux--Euler transformation depends on a 
parameter, which appears in the factorization~(3), 
that is, on the choice of a solution of the Riccati 
equation~(2). We denote these transformations by~$B$,
$$
B\:L\longmapsto\widetilde L=-(\partial_x-v)
(\partial_x+v),\qquad
\psi\longmapsto\widetilde\psi=(\partial_x-v)\psi.
\tag 7
$$

Moreover, there is still a one-parameter family of 
transformations of `energy shift'
$$
T_\alpha\:L\longmapsto L+\alpha,\qquad \alpha=\const.
$$
Let $B_\alpha=B\circ T_\alpha$.

In the theory of solitons, by means of the 
transformations $B_\alpha$ we could create well-known 
non-reflecting potentials from the zero potential. 
By one application of the transformation $B_\alpha$ 
we can obtain `soliton potentials' from the free 
operator $L_0=-\partial_x^2$, starting with 
$v=k\tanh k(x-x_0)$, $\varphi=\cosh k(x-x_0)$, 
$\alpha=k^2$.\linebreak
 By multiple application of the transformations 
 $B_{\alpha_j}$ we can obtain all rapidly\linebreak 
 decreasing non-reflecting potentials, taking each 
 time a non-vanishing solution $\varphi_j$\linebreak
`below the spectrum':
$$
u_{N+1}(x)=B_{\alpha_N}\circ B_{\alpha_{N-1}}\circ
\dots\circ B_{\alpha_0}(0),
\tag 8
$$
$$
u_0(x)=0,\ u_1(x)=-2k^2/\cosh^2k(x-x_0),\ \dotsc.
$$

In the literature on the theory of solitons 
(see~\cite{3}--\cite{5}), by means of B\"acklund--Darboux 
transformations the so-called `many-soliton potentials 
on the background of finite-zone potentials' were 
obtained by applying the transformations $B_\alpha$ 
with suitable parameters to the known periodic and 
quasiperiodic finite-zone potentials. This class of 
potentials corresponds to the limits of finite-zone 
potentials for various degenerate Riemann surfaces. 
For a long time nobody succeeded in establishing the 
relation between the class of non-degenerate periodic 
(quasiperiodic) finite-zone potentials (one-dimensional 
Schr\"odinger operators) and the theory of 
B\"acklund--Darboux transformations. We recall 
that finite-zone potentials correspond to orbits 
of the infinite-dimensional continuous commutative 
group of isospectral symmetries, the so-called 
higher analogues of~KdV, which have finite dimension 
(\cite{2},~\cite{6}--\cite{11}). These operators have 
remarkable spectral and algebro-geometric properties: 
their coefficients are calculated by means of 
theta-functions of hyperelliptic Riemann surfaces, 
and they generate remarkable fully integrable 
Hamiltonian systems.

An important idea on the connection between 
B\"acklund--Darboux transformations and the theory 
of finite-zone potentials was proposed in 
1986~(\cite{12}). In this work the author considered 
`cyclic chains' of the B\"acklund--Darboux 
transformations of length $N+1$, that is, equations 
of the form
$$
u=B_{\alpha_N}\circ B_{\alpha_{N-1}}\circ\dots\circ 
B_{\alpha_0}(u)
$$
under the condition $\alpha_N=\alpha_{N-1}=\dots=
\alpha_0=0$. Developing very interesting technical 
arguments in studying such chains as certain 
integrable non-linear systems, in~\cite{12} the 
supposition was put forward that for $N=2k$ the 
potential $u(x)$ is always a finite-zone potential 
(Weiss' hypothesis). This hypothesis was proved 
in~\cite{13} in a more general form.

Suppose that $N=2k$ and the $\alpha_j$ may 
be non-zero. If $\sum_{j=0}^N\alpha_j=0$, then the 
potential $u(x)$ in the cyclic chain of 
B\"acklund--Darboux transformations is finite-zone 
with a Riemann surface of genus no greater 
than~$N+ 1$.

If $\sum_{j=0}^N\alpha_j=\alpha\ne0$, then the 
potential $u(x)$ has an `oscillator-similar' 
asymptotic form
$$
u(x)=\frac{\alpha^2x^2}{4(N+1)^2}+O(x),\qquad |x|\to\infty.
$$
If the potential $u(x)$ is smooth and real 
\rom(without singularities\rom), then the spectrum 
of the operator $L=-\partial_x^2+u(x)$ is a 
combination of $N+1$ arithmetic progressions 
with the general difference~$\alpha$. At the same 
time there is a differential operator $A_{N+1}$ of 
order $N+1$ such that
$
[A_{N+1},L]=\alpha A_{N+1}.
$

V.~E.~Adler made a numerical investigation and 
established that the equation of cyclicity of 
the chains of transformations $B_{\alpha_j}$ 
for $\alpha\ne0$ has non-singular real solutions.
 In particular, for $N=2$ this equation is 
 transformed into the Painlev\'e-IV equation.

Operators of the cyclic chain
$$
L_0=L_{N+1},L_N,\dots,L_1,L_0,\dots
$$
have factorizations of the form
$
\alpha_j+L_j=Q_jQ_j^+
$.
>From the definition of  Darboux--Euler 
substitutions we have:
$$
\gather
L_j+\alpha_j=Q_jQ_j^+=-(\partial_x+v_j)(\partial_x-v_j),
\\
L_{j+1}=\widetilde L_j=Q_j^+Q_j.
\endgather
$$

For the operators $L_j$ we consider a set of 
`ground states' defined by the equation
$$
Q_{j-1}\psi_{0,j}=0=L_j\psi_{0,j}.
$$
We assume that $\psi_{0,j}\in\script L_2(\mathbb R)$, 
and all the translations $\alpha_j$ and the 
coefficients of the operators $Q_j$,~$Q_j^+$ are 
real. Then the sequence of `creation operators' 
gives the whole spectrum in $\script L_2(\mathbb R)$ 
for operators $L_M$ ($M\ge j$) according to the formula
$$
\gather
L_M\psi_{M-j,j}=\biggl(\sum_{K=j}^M\alpha_K\biggr)
\psi_{M-j,j},
\\
Q_M^+Q^+\dotsb Q_{j+1}^+Q_j^+\psi_{0,j}=\psi_{M-j,j},
\qquad M\ge j.
\endgather
$$
Here $L_M=L_{M+l(N+1)}$, $n=2k$, $l\in\mathbb Z$.

We note that the sequence of numbers $\sum_{K=j}^M
\alpha_K$ is a combination of finitely many 
arithmetic progressions with common difference 
$\sum_{K=0}^N\alpha_K=\alpha$, since the set of 
numbers $\alpha_j$ is periodic. The spectrum of 
the operator $L_M$ has the form $\sum_{K=j}^M
\alpha_K=\lambda_{j,M}$, $j=M,M-1,M-2,\dots$\,.

These are the basic results in the theory of 
cyclic B\"acklund--Darboux chains for the 
one-dimensional continuous Schr\"odinger 
operator~(\cite{13}).\footnote{For non-trivial 
cyclic chains of even length a solution of the 
problem of classification has not been obtained.} 
It would not be superfluous to see in the 
literature a publication containing a full 
justification of this beautiful algebraic picture 
from the viewpoint of rigorous spectral theory, 
that is, functional analysis. So far, this problem 
can be considered `more or less solved' on the 
level of the requirements of reasonable (not 
superrigorous) mathematical physics. An 
understanding of what happens and a set of rigorous 
formulae have already been obtained, but the 
rigorous completeness of the picture has not 
been proved in general.

\example{Example~1}
Let $N=0$ (a cyclic chain of period~1); we have
$$
\dotsb=L_{-1}=L_0=L_1=\dotsb,
$$
where $L_0=Q^+Q$, $\alpha+L_1=QQ^+$. Here there 
arises the Heisenberg algebra
$$
QQ^+=Q^+Q+\alpha,\qquad Q=\partial_x+\frac{\alpha x}{2}\,.
\tag 9
$$
Since $Q=\partial_x+v(x)$, we have
$$
u(x)=v_x+v^2=\frac\alpha2+\Bigl(\frac\alpha2\Bigr)^2.
$$
The eigenfunctions $\psi_{M-j,j}$ (mentioned above) 
are the eigenfunctions of the quantum oscillator
$$
L=-\partial_x^2+\frac{\alpha^2x^2}{4}
=-\Bigl(\partial_x+\frac{\alpha x}{2}\Bigr)
\Bigl(\partial_x-\frac{\alpha x}{2}\Bigr)-\frac{\alpha}{2}
=-\Bigl(\partial_x-\frac{\alpha x}{2}\Bigr)
\Bigl(\partial_x+\frac{\alpha x}{2}\Bigr)+\frac{\alpha}{2}\,.
$$
Let $\alpha>0$. The equation $Q\psi_0=0$ has a 
solution $\psi_0\in\script L_2(\mathbb R)$, $\psi_0=
e^{-\alpha x^2/4}$. All eigenfunctions 
$\psi_M=(Q^+)^M\psi_0$, $Q^+=-(\partial_x-\alpha x/2)$, 
belong to $\script L_2(\mathbb R)$ and generate the 
spectrum of the operator $L=-\partial_x^2+\alpha^2x^2/4$,
$$
\lambda_M=\frac{\alpha}{2}+M\alpha.
$$

In this example everything is clear, but we should 
establish in the general case that the 
eigenfunctions $\psi_{M-j,j}$ belong to $\script 
L_2(\mathbb R)$.
\endexample

\head
\boldsymbol\S2. The non-stationary one-dimensional 
Schr\"odinger equation
\endhead

The Darboux substitution for the non-stationary 
Schr\"odinger equation
$$
i\psi_t=-\psi_{xx}+u\psi
\tag 10
$$
is worthy of mention (see~\cite{14}). These substitutions 
have been used to construct some exact solutions of the 
KP equation. Starting from an exact solution
$$
i\varphi_t=-\varphi_{xx}+u(x,t)\varphi,
\tag 11
$$
we define the transformation
$$
B\:\psi\longmapsto\widetilde\psi=\psi_x-\frac{\varphi_x}{\varphi}\psi.
\tag 12
$$
The function $\widetilde\psi$ satisfies the equation
$$
i\widetilde\psi_t=-\widetilde\psi_{xx}+\widetilde 
u(x,t)\widetilde\psi,
\tag 13
$$
where $\widetilde u=u-2(\log\varphi)_{xx}$.

We express this in algebraic language.

\proclaim{Lemma~2}
Suppose there is given a real connection of zero 
curvature
$$
\nabla\!_t=\partial_t-w,\qquad
\nabla\!_x=\partial_x-v,\qquad
[\nabla\!_t,\nabla\!_x]=0
\tag 14
$$
such that $(10)$ has the form
$$
i\nabla\!_t\psi=\nabla\!_x^+\nabla\!_x\psi,\qquad
\nabla\!_x^+=-(\partial_x+v).
\tag 15
$$
Then the Darboux transformation has the form
$$
\widetilde\psi=\nabla\!_x\psi.
\tag 16
$$
The new function $\widetilde\psi(x,t)$ satisfies 
the equation
$$
i\nabla\!_t\widetilde\psi=\nabla\!_x\nabla\!_x^+
\widetilde\psi.
\tag 17
$$
\endproclaim

\demo{Proof}
>From (15) it follows that
$$
i\nabla\!_t\psi=\nabla\!_x^+\widetilde\psi.
$$
We apply $\nabla\!_x$ to both sides of this 
equation and use the relation of zero curvature 
$\nabla\!_x\nabla\!_t=\nabla\!_t\nabla\!_x$. We obtain
$$
i\nabla\!_x\nabla\!_t\psi
=\nabla\!_x\nabla\!_x^+\widetilde\psi=i\nabla\!_t
(\nabla\!_x\psi)
=i\nabla\!_t\widetilde\psi.
$$
We have proved Lemma~2.
\enddemo

The representation (15) is determined by one 
solution $\varphi$ of~(11), since the connection 
of zero curvature has the form
$$
\nabla\!_t=\partial_t-\frac{\varphi_t}{\varphi}\,,\qquad
\nabla\!_x=\partial_x-\frac{\varphi_x}{\varphi}\,.
\tag 18
$$
Hence, a substitution of Euler--Darboux type here 
hardly differs from the case of the stationary 
Schr\"odinger operator.

We now consider cyclic chains of length~1. Let 
the Schr\"odinger equation and a chosen solution 
$\varphi$ be such that in one step we come to 
this (gauge equivalent) equation:
$$
\gather
\widetilde u(x,t)=u-2(\log\varphi)_{xx}=u-C(t),
\\
i\varphi_t=-\varphi_{xx}+u(x,t)\varphi.
\endgather
$$
We obtain the relation
$$
\varphi(x,t)=\exp\biggl(\frac{Cx^2}{2}+Ax+B\biggr),
\tag 19
$$
where $A$, $B$, $C$ are arbitrary functions of~$t$. 
For $C\ne0$ we obtain an equation of the form
$$
\varphi(x,t)=\exp\biggl(\frac{C(x-x_0(t))^2}{2}+D(t)\biggr).
\tag 20
$$
For $C\equiv0$ we have ($A\ne0$):
$$
\varphi(x,t)=\exp\bigl(A(t)x+B\bigr)=\exp\bigl(A(t)
(x-x_0(t))\bigr).
$$

It is easy to prove the following lemma.

\proclaim{Lemma~3}
The Darboux chain for a non-stationary Schr\"odinger 
equation is cyclic of period~$1$ if and only if the 
potential $u(x,t)$ has the form
$$
u(x,t)=\bigl(\alpha(t)x\bigr)^2+\beta(t)x+\gamma(t),
$$
and the chosen solution $\varphi(x,t)$ is given in 
the form~$(19)$, where the equalities
$$
\aligned
i\dot C&=2(\alpha^2-C^2),
\\
i\dot A&=\beta-2AC,
\\
i\dot B&=\gamma-(A^2+C)
\endaligned
\tag 21
$$
are satisfied. The dynamics of $x_0$ from~$(20)$ 
is as follows\rom:
$$
\dot x_0=\frac{2\alpha^2x_0-\beta}{C}\,.
\tag 22
$$
\endproclaim

The proof is obtained by simple calculation.

We consider two cases.

\smallskip
\noindent{\sl Case}~1. Let $w+iv=C\ne0$. Then either 
$\alpha^2\ne0$ or $\alpha^2\equiv0$. In the case 
$\alpha\ne0$ we have a `moving oscillator of variable 
form'.

\proclaim{Lemma~4}
Solutions of the form~$(19)$, where $w(t_0)<0$, 
are in $\script L_2(\mathbb R)$ with respect to 
the variable $x$ for any $t\ge t_0$ if the condition
$$
\Im\alpha^2(t)\le0
$$
is satisfied.
\endproclaim

\demo{Proof}
The domain $\Re C<0$ is invariant for the system~(21), 
since for $C=w+iv$ we have $\dot w\le0$ for $w=0$ by 
virtue of the equation
$$
\dot w=2(\Im\alpha^2-2vw).
$$
We have proved Lemma~4.
\enddemo

Let $\alpha^2=\const>0$. Then the condition $C=-\alpha$, 
$\alpha\in\mathbb R$, gives a stationary solution for $C$ 
in the domain $u=\Re C<0$. All trajectories in this domain 
are periodic and the equation is easily integrated. 
The trajectories are given by the equation $H(w,v)=
\const$, where $H= ((\alpha-w)^2+v^2)/w$.

\remark{Remark~$1$}
A set of operators of the form $Q_a=\partial_x+ax$ 
satisfies the commutation relations
$$
[Q_a,Q_b]=(a-b)\cdot1,\qquad Q^+=-Q_{-\overline a}
\tag 23
$$
for all complex $a$, $b$. The commutators of all 
these operators with $H_\alpha=Q_\alpha Q_\alpha^+$, 
where $Q_\alpha^+=-Q_{-\overline\alpha}$, have the form
$$
[Q_a,H_\alpha]=(a-\alpha)Q_{-\overline\alpha}+(a+
\overline\alpha)Q_\alpha.
\tag 24
$$
The usual coherent states are eigenvectors of the 
annihilation operators. The states we have studied 
are eigenvectors of the operators
$$
Q_{-C}\psi=\gamma\psi,\qquad \Re C<0.
$$
Here $\gamma\in\mathbb C$, and in formula (20) 
above $\gamma=Cx_0$. Hence, the states (20) are 
determined purely algebraically by means of the 
Lie algebra~(23), which contains the oscillator 
Hamiltonian~$H=H_\alpha$.
\endremark

\remark{Remark~$2$}
The well-known `coherent states' of the oscillator 
are eigenvectors of the annihilation operator
$$
\gather
Q\psi_\gamma=\gamma\psi_\gamma,\qquad \gamma\in\mathbb C,
\\
Q=\partial_x+\alpha x,\qquad
\psi_\gamma(x)=e^{-\alpha(x-\gamma/\alpha)^2/2}.
\endgather
$$
Using these functions as the initial conditions, 
we obtain solutions of the \linebreak Schr\"odinger 
equation in the particular form~(20), where $C=-\alpha
=\const< 0$.\linebreak
For $x_0(t)$ we obtain motion 
along the imaginary straight line:
$$
\gather
x_0(t)=x_0(0)+2it,
\\
\psi_\gamma(x,t)=e^{-\alpha(x-
\gamma/\alpha-x_0(t))^2/2+B(t)}.
\endgather
$$
\endremark

\example{Question}
For $t=0$ suppose we have $C(0)=q$ for all solutions 
of~(20). For which $(x_{0,q},q)$ is the set of 
functions $\varphi_q(x,0)$, where
$
\varphi_q(x,0)=e^{q(x-x_{0,q})^2/2},
$
complete in $\script L_2(\mathbb R)$, where $q$ 
runs along the curve from the point $q_0=-\alpha$ 
to $q=\infty$ in such a way that $\Re q<0$? How 
can we choose a minimal complete basis if this 
set is overfull? Naturally we can choose a curve 
$\infty<q\le-\alpha$ and translations $x_{0,q}$ 
such that the set will be complete, $q\in\mathbb R$.
\endexample

\remark{Remark~$3$}
We also note that for the oscillator $u=\alpha^2x^2$, 
$\alpha^2=\const> 0$, we can perform the 
Darboux transformation, starting from the second 
solution in~$\script L_2(\mathbb R)$:
$$
\varphi(x,t)\bigr|_{t=0}=P_n(x)e^{-\alpha x^2/2},\qquad 
\alpha>0,
$$
where $P_n(x)$ is an arbitrary polynomial in $x$ 
of degree~$n$ (here $\alpha=C=\const$). For the 
potential $\widetilde u(x,t)$ we have
$$
\widetilde u=\alpha^2x^2-2(\log\varphi)_{xx}=
\alpha^2x^2-
\sum_j\frac{2}{(x-x_j(t))^2}\,.
$$
We determine the dynamics of $\varphi(x,t)$ by 
expanding $P_n(x)$ in terms of the Hermite 
polynomials, that is, in terms of eigenfunctions 
of the oscillator. There arises a motion of the 
poles of the potential $\widetilde u(x,t)$ with 
respect to~$t$ (or the motion with respect to $t$ 
of the polynomial $P_n(x,t)$, $P_n(x,0)=P_n(x)$) 
such that
$$
\gather
P_n(x,t)=\sum e^{-i\lambda_jt}a_jH_j(x)
\\
P_n(x)=\sum_{j\le n}a_jH_j,\quad\lambda_j=\alpha+2j\alpha.
\endgather
$$
$H_j(x)$ are the Hermite polynomials defined by 
the formulae
$$
H_0=1,\qquad H_j(x)=\bigl((Q^+)^je^{-\alpha x^2}\bigr)
/e^{-\alpha x^2},
$$
where $Q^+=\partial_x-\alpha x$. We have
$$
P_n(x,t)=a_ne^{-i(2n+1)\alpha t}\prod_{j=1}^n
\bigl(x-x_j(t)\bigr).
$$

We can obtain a more extensive family of solutions 
of this type from initial conditions of the form
$$
\varphi(x,0)=P_n(x)e^{C(0)(x-x_0)^2/2},
$$
applying the operators $Q_{-C}^+=Q_{\overline C}$ 
to the initial conditions for solutions of~(20). 
The dynamics of their poles can also be of interest.
\endremark

\smallskip
\noindent{\sl Case}~2. Let $C\equiv0$. Then 
$\alpha\equiv0$ and we have 
$$
u(x,t)=\beta x+\gamma,\qquad
\varphi(x,t)=\exp\bigl(iA(t)(x-x_0(t))\bigr).
$$
This is a physically reasonable case when $\beta=E(t)$ 
and is periodic in~$t$. In this case (known as the 
integrable case of L.~V.~Keldysh) we can also find 
a general solution.

We note with interest that even in the stationary 
case $u(x)=\beta x+ \gamma$ when 
$\beta,\gamma$ are constants, it is more convenient 
to use the basis of non-stationary solutions 
$\varphi(x,t)$, which have the form of plane 
waves for fixed time.

Ultimately only the non-stationary Schr\"odinger 
equation is a law of nature. If the force $u_x$ 
does not depend on time, then we can consider the 
Hamiltonian $\widehat H$ in an appropriate gauge 
as stationary. As usual, in this case in place of 
the non-stationary equation
$$
i\psi_t=\widehat H\psi
$$
we can use the Fourier method and solve the 
stationary problem
$$
\widehat H\psi=\lambda\psi
$$
in the space $\script L_2(\mathbb R)$.

After this, we shall have the general solution 
in the form of finite or continuous linear combinations
$$
\psi(x,t)=\sum_ja_j\psi_je^{-i\lambda_jt},\qquad
\widehat H\psi_j=\lambda_j\psi_j.
$$
However, this generally accepted approach makes 
sense only if the spectral problem is sufficiently 
well solved: it is only a mathematical trick, nothing 
more.

If the stationary problem is not sufficiently 
well solved, then the Fourier method  is not 
appropriate. An example of such a situation is 
the constant electric field $u(x)=Ex$, where the 
basis of non-stationary solutions (plane waves) 
is much simpler, as we mentioned above, than the 
eigenfunctions of the stationary operator~$\widehat 
H$ (Airey functions).

The advantage of this approach can be particularly 
important in the case when the potential has the 
form
$$
u(x)=Ex+u_0(x),
$$
where the function $u_0(x)$ is periodic (a constant 
electric field is applied to the crystal). Already 
in~\cite{15} the perspectives of the non-stationary 
approach in this case were discussed. A~physically 
reasonable general non-stationary case is
$$
u(x,t)=E(x,t),
$$
where $E(x,t)$ is the electric field, periodic in 
both variables. If the potential $u(x,t)$ is doubly 
periodic (`zero electric current'), then we have an 
extensive family of algebro-geometric exactly soluble 
equations, where $u(x,t)$ is expressed in terms of 
theta-functions of Riemann surfaces 
(see~\cite{16},~\cite{17}). We construct the Bloch 
solution $\psi$ of~(10):
$$
\aligned
\psi(x+T_1,t)&=e^{ip_1T_1}\psi(x,t),
\\
\psi(x,t+T_0)&=e^{ip_0T_0}\psi(x,t),
\endaligned
\tag 25
$$
as the `Baker--Akhiezer function' on some Riemann 
surface $\Gamma$ of finite genus, that is, the 
parameters $p_1$,~$p_0$ are connected by the equation 
of the algebraic curve~$\Gamma$. These potentials 
were used for solutions of the `KP~equation', where 
$t$ is renamed as~$y$, and time, denoted by~$t$, 
is the third variable. Undoubtedly these solutions 
of Krichever~\cite{16},~\cite{17} are obtained from 
the cyclic chains of Darboux transformations, but a 
rigorous assertion has not been proved up to now.

More complex is the case of a non-zero `electric flux'. 
The case $E(t)=\beta$ is well known as the integrable 
case of Keldysh.

In these cases, to understand the problem it would be 
useful to broaden the number of known exact solutions. 
In~\cite{15} under certain conditions (`quantization of 
the electric flux' using an elementary cell in the 
$(x,t)$ plane) Novikov introduced electric Bloch 
functions, which are eigenfunctions for `electric 
translations' by periods, but their analytical 
properties for $E\ne0$ have not been understood and 
formulated up to now.

We transform the Schr\"odinger equation (10) using 
the gauge transformation
$$
\psi'=e^{iF(x,t)}\psi,\qquad L'=e^{iF}Le^{-iF},
\tag 26
$$
so that
$$
i\psi'_t=-\bigl(\partial_x+iA_1(x,t)\bigr)^2\psi'.
$$
For this we need to take $F_t=u(x,t)$, $A_1=-F_x$.

Let $u=\beta(t)x$. We have $F=A(t)$, $\dot A=\beta(t)$. 
We have a solution of form
$$
\psi'_k=e^{ikx}\varphi_k(t),
\tag 27
$$
where
$$
i\dot\varphi_k=\bigl(k-A(t)\bigr)^2\varphi_k,\qquad
\varphi_k=e^{-i\int^t(k-A(\tau))^2}d\tau.
\tag 28
$$
For
$$
\beta(t)=E=\const
\tag 29
$$
we have
$$
\psi_k'=e^{ikx}e^{iE^2/3}(kE^{-1}-t)^3.
\tag 30
$$
This is a basis in the space of solutions.

The operators of `electric translations', which 
preserve the Schr\"odinger equation,
$$
i\nabla\!_t\psi=-\nabla\!_x^2\psi,
\tag 31
$$
where $\nabla\!_t=\partial_t+iA_0$, $\nabla\!_x=
\partial_x+iA_1$, are fully analogous to magnetic 
translations. For a periodic force $E(x,t)$,
$$
E(x,t)=A_{0x}-A_{1t}=E(x+T_1,t)=E(x,t+T_0),
$$
the electric translations that commute with (31) 
are given by
$$
\widehat T_1\psi=e^{iF_0(x,t)}\psi(x+T_1,t),\qquad
\widehat T_0\psi=e^{iF_1(x,t)}\psi(x,t+T_0),
\tag 32
$$
where $dF_0=A(t+T_0)-A(t)$, $dF_1=A(x+T_1)-A(x)$, 
$A=A_0\,dt+A_1\,dx$. We have the identity
$$
\widehat T_1\widehat T_0-\widehat T_0\widehat 
T_1=e^{i\Phi_{01}},\qquad
\Phi_{01}=\int_0^{T_0}\int_0^{T_1}E(x,t)\,dx\,dt.
\tag 33
$$
If $\Phi_{01}=2\pi q$, $q\in\mathbb Z$, then the 
group of `electric translations' is commutative 
(as in the magnetic case). In the particular case 
(29) we obtain
$$
\widehat T_1\psi_k'=T_1\psi_k'=e^{ikT_1}\psi_k',\qquad
\widehat T_0\psi_k'=\psi_{k-ET_0}'.
\tag 34
$$
The electric Bloch states, which are eigenstates 
for $\widehat T_1,\widehat T_0$ under the condition 
$ET_0T_1=2\pi q$, $q\in\mathbb Z$, are formally 
written in the form
$$
\psi(x,t,p_0,p_1)=\sum_{m\in\mathbb Z}e^{-imp_0T_0}
\widehat T_0^m\bigl(\psi_k'(x,t)\bigr).
\tag 35
$$
Obviously we have
$$
\gather
\widehat T_0\psi=e^{ip_0T_0}\psi,
\\
\widehat T_1\psi=e^{ip_1T_1}\psi,\qquad
p_1=k\pmod{ET_0}.
\endgather
$$
These are complicated generalized functions. For 
$\beta(t)=E$ we have, for example,
$$
\psi(x,t,p_0,p_1)=e^{ikx}\biggl(\sum_{m\in\mathbb Z}
e^{i(-mT_0(p_0+Ex)+\frac{E^2}{3}(kE^{-1}-t-mT_0)^3)}
\biggr).
\tag 36
$$
(As opposed to the magnetic case, where the 
functions $\varphi_k$ rapidly decrease as 
$|x|\to\infty$, and the analogue of the series (35) 
converges, giving analytic functions.)

As the simplest example we show that under the 
`strong integer' condition 
$q,s,l\in\mathbb Z$
$$
ET_0T_1=2\pi q,\qquad E^2/3=2\pi s,\qquad T_0=l,
\tag 37
$$
we obtain for $t_0=kE^{-1}$, $0\le k\le ET_0$
$$
\frac{1}{2\pi}\psi(x,t_0,p_0,p_1)=
e^{ikx}\sum_{m\in\mathbb Z}\delta
\bigl(T_0(p_0+Ex)+2\pi m\bigr),\quad k=p_1\pmod{ET_0}.
\tag 38
$$

Hence, under the condition (37) the electric 
Bloch states are obtained by solving the Cauchy 
problem with singular initial condition~(38). 
In the survey~\cite{15}, although the scheme 
and the ideology were correct, misprints and 
errors were made in the formulae of the text 
in the description of the electric Bloch states.

The function (36) is a sum of two functions
$
\psi=\psi_++\psi_-,
$
where $\psi_+$ is the boundary value of the 
function which is analytic for $\Im p_0> 
0$, and $\psi_-$ is the boundary value of the function 
which is analytic for $\Im p_0< 0$. 
This property is also possibly true in the general 
case for electric Bloch states. Our series (36) 
converges for $p_1=k_{\mathbb R}+i\kappa^2$, 
$\kappa^2>0$, as we can easily see from~(36), 
so in the variable $k=p_1\pmod{ET_0}$ the function 
(36) is the boundary value of an analytic function.

In the case of a constant field under the strong 
integer condition~(37), starting from (38) the 
electric Bloch function can be completely calculated. 
We recall that a theta-function with zero 
characteristics has the form
$$
\theta_\tau(z)=\sum_{m\in\mathbb Z}e^{i\tau m^2/2+mz}
\tag 39
$$
and is well defined for $\Im\tau>0$.

Comparing with (36) we have finally
$$
\psi(x,t,p_0,p_1)=\theta_\tau(z),\quad
\tau=2(ET_0)^2\Bigl(\frac{k}{E}-t\Bigr)^2+i0,\quad
z=i(ET_0)\Bigl(k-Et-\frac{p_0}{E}\Bigr),
\tag 40
$$
that is, the limit of the theta function 
when $\Im\tau\to+0$,
$$
\gather
p_1=k\pmod{ET_0},
\\
\widehat T_1\psi=e^{ip_0T_0}\psi,\qquad
\widehat T_2\psi=e^{ip_1T_1}\psi,
\\
T_0\in\mathbb Z,\quad E^2/3\in2\pi\mathbb Z,
\quad ET_0T_1\in2\pi\mathbb Z.
\endgather
$$
In the remaining cases there arise generalized
functions (distributions) given by infinite 
trigonometric sums that are cubic in 
$m\in\mathbb Z$. Apparently, quantities of 
this type have not been studied.

\remark{Remark~$4$}
In the case $\alpha=0$ and $\beta=\const=E$ we 
have a special family of solutions such that 
$C\ne0$ for $t\in\mathbb R$. From (21) we find 
that $i/(2t+ip)=C$, $p\in\mathbb R$,
$$
\varphi(x,t)=\exp\biggl\{i\frac{(x-x_0(t))^2}{4t+2ip}+
D(t)\biggr\}.
$$
We have
$$
-x_0(0)+x_0(t)=\beta t^2,
$$
where $p<0$.

The motion of the centre $x_0$ for $p=0$ is the 
motion of a classical particle in the constant 
field $E=\beta$. In this solution we have coincidence 
of the classical and quantum pictures.
\endremark

We have presented these elementary solutions here 
to demonstrate on the example $N=1$ how the 
problem on cyclic Darboux chains for the one-dimensional 
non-stationary Schr\"odinger operator should be 
correctly formulated.

\definition{Definition~2}

We call the transformation
$$
u\to u+\gamma(t),\qquad \psi\to e^{-ig(t)}
\tag 41
$$
the {\it gauge transformation} of a 
one-dimensional non-stationary Schr\"odinger operator, 
where $\gamma(t)$ is an arbitrary function of time, 
and $\dot g=\gamma$.
\enddefinition

\definition{Definition~3}
The {\it chain of Darboux transformations} for a 
one-dimensional non-stationary Schr\"odinger 
operator is a sequence of equations
$$
i\psi_t=L_j\psi=-\psi_{xx}+u_j(x,t)\psi
$$
such that the operator $L_{j+1}$ is obtained 
from $L_j$ by the Darboux transformation and 
then by the gauge transformation acting on the 
function~$\gamma_j(t)$:
$$
\gathered
u_{j+1}(x,t)=\widetilde u_j(x,t)+\gamma_j(t),
\\
\widetilde u_i=u_i-2(\log\varphi_i)_{xx},
\\
i\varphi_{jt}=-\varphi_{jxx}+u_j(x,t)\varphi.
\endgathered
\tag 42
$$
\enddefinition

Hence, the Darboux chain is determined by a 
choice of solutions $\varphi_j$ and arbitrary 
functions of time $\gamma_j(t)$, $j\in\mathbb Z$. 
In the case of a cyclic chain of period $N$ we 
have $N$ solutions $\varphi_j$ and $N$ arbitrary 
functions~$\gamma_j(t)$.

\head
\boldsymbol\S3. One-dimensional difference operators
\endhead

We consider a difference operator $L$ of the second order
$$
L\psi_n=c_{n-1}\psi_{n-1}+v_n\psi_n+c_n\psi_{n+1}.
\tag 43
$$
>From the theory of solitons we know~\cite{3}, 
\cite{18}--\cite{20}, that for the conservation, 
in the discrete case, of the latent algebraic 
symmetries of Schr\"odinger operators (such as 
isospectral deformations of KdV type and the 
B\"acklund--Darboux--Euler transformation) it 
is necessary to introduce the covariant 
translations $\exp(\nabla\!_x)$ in place 
of the translations $T=\exp(\partial_x)$, where
$
\exp(\nabla\!_x)=c_nT\:\psi_n\to\psi_{n+1}c_n
$
(the `covariant translation').

After this, the theory of the corresponding 
class of difference operators (43) is similar 
to the continuous case: there arise isospectral 
deformations of the type of the Toda chain for 
(43) or of the `discrete~KdV' for (43) under 
the condition $v_n=0$. There also arise analogues 
of Darboux transformations starting from factorization:
\roster
\Item"a)" $\alpha+L=QQ^+$, $T^+=T^{-1}$, 
$Q=a_n+b_nT$, $Q^+=a_n+T^{-1}b_n$;
\Item"b)" $\alpha+L=\widehat Q^+\widehat Q$,
$\widehat Q=\widehat a_n+\widehat b_nT$.
\endroster

For the factorization, as before, we have 
to solve a difference analogue of the Riccati 
equation
$$
v_n+\alpha=a_n^2+b_n^2,\qquad a_{n+1}b_n=c_n,
\tag 44
$$
$$
v_n+\alpha=\widehat a_n^2+\widehat b_{n-1}^2,\qquad
c_n=\widehat a_n\widehat b_n.
\tag 45
$$

Factorization is also possible in the general 
non-self-adjoint case (direct and inverse):

$$
L=p_nT^{-1}+q_n+r_nT=(a_n+b_nT)(x_n+y_nT^{-1})=
(c_n+d_nT^{-1})(v_n+w_nT).
$$
 
The B\"acklund--Darboux transformations are 
defined as in the continuous case (\cite{21},~\cite{22}):
$$
B_\alpha=BT_\alpha,\qquad B_\alpha'=B'T_\alpha,
\tag 46
$$
$$\begin{aligned}
B&\:L\longmapsto\widetilde L=Q^+Q,&&\quad
\psi\longmapsto\widetilde\psi=Q^+\psi,
\\
B'&\:L\longmapsto\widehat L'=\widehat Q\widehat Q^+,&&\quad
\psi\longmapsto\widehat\psi'=\widehat Q\psi,
\end{aligned}
\tag 47
$$
$$T_\alpha\:L\longmapsto L+\alpha,
\tag 48
$$
$L=QQ^+$ (in all cases we first perform factorization 
and then permute the non-commutative factors, that is, 
the first-order operators).

There naturally arises the problem on cyclic 
chains (\cite{21},~\cite{22}), for example, on 
the direct transformations
$$
B_{\alpha_N}\circ B_{\alpha_{N-1}}\circ\dots\circ 
B_{\alpha_0}(L)=L.
\tag 49
$$
It was established in~\cite{22} that if 
$\sum\alpha_j=0$, then $L$ is a finite-zone 
operator with Riemann surface (`spectrum') of 
genus $g\le[N/2]$. If the $\sum\alpha_j=\alpha\ne0$, 
then we arrive at the difference analogues of 
the theory~\cite{12},~\cite{13}.

We should take into account that the 
transformations $B_\alpha$ and $B_\alpha'$ depend 
also on a continuous parameter, that is, on the 
choice of a solution of the Riccati difference 
equation (44) or~(45), which depends on the choice 
of the initial point~$b_0$ (or~$\widehat b_0$) 
and the set of signs ($\sgn a_n$). We denote this 
joint parameter by~$\kappa$. The problem of an 
appropriate choice of signs is discussed, for 
example, in~\cite{21}. In any case, we have the relation
$$
B^{\prime\kappa'}B^\kappa L=L
$$
for any $L$, where the choice of the parameter 
$\kappa'$ is naturally consistent with~$\kappa$.

In principle we can consider general words in 
the group generated by the transformations 
$T_{\alpha'}B^{\prime\kappa}$, $B^\kappa$ for 
all $(\kappa,\alpha)$ and formulate the cyclicity 
condition with respect to them. Hence, we have 
the following conclusion.

\smallskip\noindent{\bf Conclusion.}
The set of Darboux transformations for a 
one-dimensional difference operator is larger 
than for a continuous one because of the two 
different types of factorizations (above). 
In~\cite{21}--\cite{23} the cyclicity condition 
was studied for `direct' chains of the form (49) only.

It is interesting to look at the case $N=1$, 
where we obtain an analogue of the oscillator, 
starting from the Heisenberg relation:
$$
QQ^+=Q^+Q+\alpha.
\tag 50
$$
This equation has a solution
$$
Q^+=1+\sqrt{a+bn}\,T.
\tag 51
$$
The coefficients of such an operator $Q$ 
cannot be real for all~$n$.

\proclaim{Lemma~5}
We assume that the `quantization condition'
$$
a+bn=b(n_0+n)
\tag 52
$$
is satisfied, where $n_0=a/b\in\mathbb Z$, 
$a,b\in\mathbb R$, $b>0$.

Then the operators $Q$ and $Q^+$ are well 
defined in the subspace of functions $\psi$ 
such that
$$
\psi_n=0,\qquad n+n_0\le0,\qquad \psi\in\script 
L_2(\mathbb R).
$$
\endproclaim

This lemma is easily proved by direct verification.

As was shown in~\cite{22}, the ground state 
is defined according to Dirac:
$$
Q^+\psi_0=0.
$$
At the same time we have~\cite{22}
$$
\psi_{0n}^2=1/(\alpha^{k-1}(k-1)!),\qquad k=n+n_0>0.
\tag 53
$$
Hence, $\psi_{0n}^2$ is the Poisson distribution. 
Applying creation operators we obtain the eigenfunctions
$$
Q^k\psi_0=P_k(n+n_0)\psi_{0n},
$$
where $P_k(n)$ are the well-known Charlier 
polynomials, orthogonal with respect to the 
weight $(\psi_{0n}^2)$, that is, the Poisson 
distribution on~$\mathbb Z_+$.

Hence, in the difference case the Heisenberg 
relation (50) is realized (among first-order 
operators) on the positive half-line~$\mathbb Z_+$.

There is also another analogue, the `$q$-analogue' 
of the oscillator, where the operators $Q$,~$Q^+$ 
depend on two parameters:
$$
Q_{c,a}=1+ca^nT.
\tag 54
$$

\proclaim{Lemma~6}
We have the relation
$$
Q_{c,a}Q_{c,a}^+-1=a^2(Q_{c_1,a}^+Q_{c_1,a}-1),\qquad 
c_1=ca^2.
\tag 55
$$
\endproclaim

The proof consists of direct verification.

In \cite{22} the following theorem was proved 
(parts of it were already given in~\cite{21}).

\proclaim{Theorem~1}
$1)$ The transformation
$$
\tau\:n\to1-n
$$
acts on the operators $Q_c$, $Q_c^+$ in the 
following way\rom:
$$
\tau Q_{c,a}=Q^+_{c,a^{-1}}\tau.
\tag 56
$$

$2)$ The equation $Q_c\psi_0=0$ has a solution 
in $\script L_2(\mathbb Z)$ under the condition 
$|a|>1$. The equation $Q_c^+\psi_0=0$ has a 
solution in $\script L_2(\mathbb Z)$ under 
the condition~$|a|<1$.

$3)$ The spectrum of the operator 
$L_{c,a}=Q_{c,a}Q_{c,a}^+$ in the interval 
$0\le\lambda<1$ has the form
$$
\begin{aligned}
&\lambda_n=1-a^{-2n},&&\quad n\ge0,
\quad\text{if}\quad |a|>1,
\\
&\lambda_n=1-a^{2n},&&\quad n\ge1,\quad\text{if}
\quad |a|<1.
\end{aligned}
\tag 57
$$
Analogously, the spectrum of the operator 
$\widetilde L_{c,a}=Q^+_{c,a}Q_{c,a}$ in the 
interval $0\le\lambda<1$ is given by
$$
\begin{aligned}
&\lambda_n=1-a^{-2n},&&\quad n\ge1,\quad 
\text{if}\quad |a|>1,
\\
&\lambda_n=1-a^{2n},&&\quad n\ge0,\quad 
\text{if}\quad |a|<1.
\end{aligned}
\tag 58
$$
\endproclaim

\remark{Problem}
Investigate the spectrum of the operators 
$L=QQ^+$, $\widetilde L=Q^+Q$ in $\script 
L_2(\mathbb Z)$ in the domain $\lambda\ge1$, 
if $Q$ has the form~(54). According to the 
hypothesis~\cite{22}, this spectrum is continuous 
and Lebesgue (see \S6 below).
\endremark

\head
\boldsymbol\S4. The Laplace transformation 
and the two-dimensional Schr\"odinger operator 
in a magnetic field
\endhead

We write a general $n$-dimensional stationary 
scalar Schr\"odinger operator in Euclidean 
space in the form
$$
L=-{\frac12}\sum\bigl(\partial_\alpha+iA_\alpha(x)
\bigr)^2+U(x),
\tag 59
$$
where $x=(x^1,\dots,x^n)$, $\partial_\alpha=
\partial/\partial x^\alpha$. In particular, 
the 2-form of the `magnetic field' has the form
$$
\gathered
H=\sum_{\alpha<\beta}H_{\alpha\beta}dx^\alpha
\wedge dx^\beta,\qquad
dH=0,
\\
H_{\alpha\beta}=\partial_\alpha A_\beta-
\partial_\beta A_\alpha
=-H_{\alpha\beta}.
\endgathered
\tag 60
$$
The electric force has the form
$$
\vec E=(\partial_1U,\dots,\partial_nU),
\tag 61
$$
that is, the potential $U$ is defined up to a 
constant. By the gauge transformations
$$
L\longmapsto e^fLe^{-f},\qquad \psi\longmapsto e^f\psi
\tag 62
$$
we can change the `vector-potential':
$$
A_\alpha\longmapsto A_\alpha-i\partial_\alpha f.
\tag 63
$$

In the physically comprehensible self-adjoint 
case we have a real magnetic field and the 
potential: $H_{\alpha\beta}\in\mathbb R$, 
$U\in\mathbb R$, and we choose a real vector-potential 
$A_\alpha\in\mathbb R$. Then the gauge 
transformations have the form (62) (in the class 
of stationary operators), where $f=i\varphi(x)$, 
$\varphi(x)\in\mathbb R$. The operator obtained 
by this transformation possesses as before the 
formal self-adjoint property with respect to the 
standard scalar product in the Hilbert space~$\script 
L_2(\mathbb R)$:
$$
\langle\psi_1,\psi_2\rangle=\int_{\mathbb R^n}
\psi_1\overline\psi_2\,d^nx.
\tag 64
$$

An important class consists of operators with 
smooth real magnetic and electric fields, which 
are periodic with respect to some lattice 
$\Gamma\subset\mathbb R^n$ of rank~$n$. In 
this case we introduce the following notions.

\definition{Definition~4}
We say that an operator $L$ is {\it topologically 
non-trivial} if the magnetic field (60) is the 
non-zero cohomology class of a torus,
$$
0\ne[H]\in H^2(\mathbb T^n,\mathbb R).
\tag 65
$$
\enddefinition

\definition{Definition~5}
We say that an operator $L$ {\it possesses a 
topologically non-trivial electric field} if 
the 1-form of the electric field (in the 
non-relativistic stationary formalism) is the 
non-zero cohomology class of a torus:
$$
0\ne[E]\in H^1(\mathbb T^n,\mathbb R),
\tag 66
$$
$E=E_\alpha dx^\alpha$, $E_\alpha=\partial_\alpha 
U(x)$. In other words, the potential has the form
$$
U(x)=E_{0\alpha}x^\alpha+U_0(x),
\tag 67
$$
where the function $U_0(x)$ is periodic with respect 
to the lattice, $E_0\!=\!\const\!\in\!(\mathbb 
R^n\setminus 0)$.
\enddefinition

A topologically non-trivial electric field has 
arisen already in the one-dimensional case. As 
we showed in~\S2, it was appropriate to study 
such a situation in the framework of the 
non-stationary $(x,t)$-formalism, even if the 
electric force is stationary.

For magnetic fields we use the stationary formalism. 
If $\gamma_1,\dots,\gamma_n\in\Gamma$ is the basis 
of the lattice, then $H_{\alpha\beta}(x+\vec\gamma_j)=
H_{\alpha\beta}(x)$. For the vector-potential we have
$$
A_\alpha(x+\vec\gamma_j)=A_\alpha(x)+\partial_\alpha
\varphi_j(x),
$$
where $\varphi_j$ is some function of $x=(x^1,\dots,x^n)$.

We define `magnetic translations' (\cite{15}), 
which have been well known in the physical literature 
since the mid-60s:
$$
\widehat T_j\psi(x)=\psi(x+\vec\gamma_j)e^{i\varphi_j(x)}.
\tag 68
$$
We have the relation
$$
\widehat T_j\widehat T_k=e^{i\Phi_{jk}}\widehat 
T_k\widehat T_j,
\tag 69
$$
where $\Phi_{jk}$ is the flux of a magnetic field 
through an elementary two-dimensional cell of the 
lattice in the plane $(\vec\gamma_j,\vec\gamma_k)$, 
or the scalar product (integral) of the cocycle 
$[H]$ with the basis 2-cycle $Z_{jk}\in H_2(\mathbb 
T^n,\mathbb Z)$.

In the two-dimensional case $n=2$ we can present 
the operator $L$ in a factorized form, which we 
write as
$$
2L=(\overline\partial+B)(\partial+A)+2W,
\tag 70
$$
where $A(z,\overline z)$, $B(z,\overline z)$, 
$W(z,\overline z)$ are some functions of 
$z$,~$\overline z$,
$$
z=x+iy,\qquad\partial=\partial_x-i\partial_y.
$$
The magnetic field has the form
$$
2H=B_z-A_{\overline z}.
\tag 71
$$

If $H$, $W$ are real, then we can, by using the 
gauge transformation, bring the operator to the 
self-adjoint form, so we have
$$
\gather
2L=Q_1Q_2+2W,
\\
Q_1=\overline\partial+B,\qquad
Q_2=\partial+A=-Q_1^+,\qquad
A=-\overline B.
\endgather
$$
The operator $L$ is always written in the form~(70), 
that is, it `admits factorization' in our terminology.

The operator $L$ also admits an opposite 
factorization of the form 
$$
2L=Q_2Q_1+2V,
$$
where $V=W+H$, $Q_2=\partial+A$, $Q_1=\overline\partial+B$.

\definition{Definition~6}
The {\it Laplace transformation} is the transformation
$$
L\longmapsto\widetilde L,\qquad
2\widetilde L=WQ_2W^{-1}Q_1+2W,
\tag 72
$$
where $2L=Q_1Q_2+2W$.
\enddefinition

\proclaim{Lemma~7}
\rom{a)} If $\psi$ is an arbitrary solution of 
the equation $L\psi=0$, then the function 
$\widetilde\psi=Q_2\psi$ satisfies the equation 
$\widetilde L\widetilde\psi=0$.

\rom{b)} If $W=W_0=\const$, then for any solution 
$\psi$ of the equation $L\psi=\lambda\psi$ the 
function $\widetilde\psi=Q_2\psi$ satisfies the 
equation $\widetilde L\widetilde\psi=
\lambda\widetilde\psi$. In this case $2\widetilde 
L=Q_2Q_1+2W_0$.
\endproclaim

\demo{Proof}
The proof of Lemma~7 is very simple. Since $L\psi=0$, 
we have
$$
-2W\psi=Q_1Q_2\psi=Q_1\widetilde\psi.
$$
>From this it follows that
$$
\gather
-2\psi=W^{-1}Q_1\widetilde\psi,
\\
-2\widetilde\psi=-2Q_2\psi=Q_2W^{-1}Q_1\widetilde\psi,
\\
-2W\widetilde\psi=WQ_2W^{-1}Q_1\widetilde\psi.
\endgather
$$
The lemma has been proved for the case~a). In the 
case~b), for $W=W_0=\const$, the proof is exactly 
the same.
\enddemo

\proclaim{Lemma~8}
The Laplace transformation for the equation $L\psi=0$ 
commutes with the gauge transformation~$(62)$. If 
$W=W_0=\const$, then this applies to the whole 
family of equations $L\psi=\lambda\psi$.

In particular, this transformation is defined on 
the gauge invariant quantities $H$,~$W$\rom:
$$
\widetilde H=H+{\frac12}\partial\overline
\partial\log W,\qquad
\widetilde W=W+\widetilde H.
\tag 73
$$
\endproclaim

The proof of this lemma is obtained by direct 
verification. We omit it.

\remark{Remark~$5$}
The classicists, beginning with Laplace, 
considered this transformation for the hyperbolic equation

1) $2L=(\partial_x+A)(\partial_y+B)+2W$,
$$
2\widetilde L=W(\partial_y+B)W^{-1}(\partial_x+A)+2W;
$$

2) $2L=(\partial_y+B)(\partial_x+A)+2V,\quad V=W+H$,
$$
2\widetilde L'=V(\partial_x+A)V^{-1}(\partial_y+B)+2V.
$$

Here the pair of mutually inverse Laplace 
transformations acts on solutions of the 
equation~$L\psi=0$:

1) $L\longmapsto\widetilde L$,\quad
$\psi\longmapsto\widetilde\psi=(\partial_y+B)\psi$;

2) $L\longmapsto\widetilde L'$,\quad
$\psi\longmapsto\widetilde\psi'=(\partial_x+A)\psi$.

As before, the gauge invariants have the form 
$H$,~$W$ or $H$,~$V$, where
$$
2H=\partial_xB-\partial_yA.
$$
For these invariants the Laplace transformations 
have the form analogous to~(73):
$$
\aligned
1)&\quad\widetilde H=H+{\frac12}\partial_1\partial_2
\log W,\quad
\widetilde W=W+\widetilde H;
\\
2)&\quad\widetilde H'=H+{\frac12}\partial_1\partial_2
\log V,\quad
\widetilde V'=V+\widetilde H'.
\endaligned
\tag 74
$$
\endremark

These transformations have been studied 
geometrically; they have found applications 
in the theory of congruences of surfaces in~$\mathbb 
R^3$~(\cite{24}). The classicists of geometry 
(Darboux, Tzitz\'eica, Moutard, 
see~\cite{25},~\cite{26}) studied chains of Laplace 
transformations and carried out a series of 
useful formal calculations. They observed that 
the infinite chain of Laplace transformations
$$
\dots,L_{-1},L_0,L_1,L_2,\dots,
$$
where $L_{j+1}=\widetilde L_j$, is equivalent to 
a non-linear evolution system, which in the 
modern literature is called the `two-dimensionalized 
Toda chain'. Integrability of it by the methods of 
soliton theory was discovered in~\cite{27}, and 
from the viewpoint of field theory and Lie algebras 
in~\cite{28},~\cite{29}. In fact, a series of 
formal results had already been obtained at the 
beginning of the 20th century.

We consider the following quantities, that is, 
the potentials of a chain:
$$
W_n=\exp f_n.
$$
>From (74) we obtain
$$
W_{n+1}-W_n=e^{f_{n+1}}-e^{f_n}=H_{n+1},\qquad
H_{n+1}-H_n={\frac12}\partial_x\partial_y(f_n).
$$
By means of the substitution $f_n=g_n-g_{n-1}$ 
we obtain the `two-dimensionalized Toda chain'
$$
{\frac12}\square g_n=e^{g_{n+1}-g_n}-e^{g_n-g_{n-1}},
\tag 75
$$
where $\square=\partial_x\partial_y$.

Darboux posed the problem of classifying the 
cyclic Laplace chains $L_N=L_0$ of period~$N$. 
For $N=2$ he deduced, in particular, that 
solutions of the equation
$$
\square g=\sinh g
\tag 76
$$
define the class of cyclic chains.

His student Tzitz\'eica (\cite{26}) posed the 
problem: let the operator $L_0= L_N$ 
be such that $H_0=H_N=0$. What kind of non-linear 
systems can arise in this case? For $N=3$ he 
showed that such cyclic chains are generated, 
in particular, by solutions of the equation
$$
\square g=e^g-e^{-2g}.
$$
However, in the hyperbolic case, which was 
considered by the classicists, there are nothing 
but formal identities.

Well posed global problems arise only in the elliptic 
case $\partial_x\to\overline\partial$, 
$\partial_y\to\partial$, which we also consider 
in relation to the spectral theory of the 
Schr\"odinger operator in electric and magnetic 
fields for $n=2$, in particular when the fields 
are smooth non-singular real doubly periodic 
functions with some lattice of periods $\Gamma$ 
on the plane~$\mathbb R^2$.

\example{Example~2}
Let $N=1$. We have
$$
\gather
L_1=L_0,\qquad H_1=H_0+{\frac12}\Delta f_0,
\\
e^{f_1}=e^{f_0}+H_1.
\endgather
$$
Since $H_1=H_0$, we obtain
$
\Delta f_0=0
$.
If $f_0$ is a non-singular doubly periodic 
function (that is, $W_0\ne0$), then $f_0= 
\const$.

Since $f_1=f_0$, we have $H_1=0=H_0$. Thus, we obtain 
only a free operator in the zero magnetic field, 
but this is a consequence of the global hypothesis.
\endexample

\example{Example~3}
Let $N=2$. We have
$$
f_2=f_0,\qquad H_2=H_0.
$$
Since $H_1=H_0+{\frac12}\Delta f_0$, 
$H_2=H_0=H_1+{\frac12}\Delta f_2$, we obtain
$$
{\frac12}\Delta(f_0+f_1)=0,
$$
or
$$
f_0=-f_1+a.
$$
Further, since
$$
\gather
e^{f_2}=e^{f_0}=e^{f_1}+H_0,
\\
e^{f_1}=e^{f_0}+H_1=e^{f_0}+H_0+{\frac12}\Delta f_0,
\endgather
$$
we obtain
$$
\gather
2(e^{f_1}-e^{f_0})={\frac12}\Delta f_0,
\\
2(e^{-f_0+a}-e^{f_0})={\frac12}\Delta f_0.
\endgather
$$
Let $g=f_0-a/2$. We obtain
$$
\gather
2(e^{-g+a/2}+e^{g+a/2})={\frac12}\Delta g,
\\
e^{a/2}\sinh g=\frac{1}{8}\Delta g.
\endgather
$$
This shows that in the non-singular doubly periodic 
case all the cyclic Laplace chains of length $N=2$ 
are described by solutions of the equation
$$
\gamma e^{a/2}\sinh g=\Delta g.
$$
By stretching the axes $(x,y)$ we reduce it to the 
sinh-Gordon equation
$$
\Delta g=\sinh g.
\tag 77
$$
\endexample

\example{Example~4}
Let $N=3$, $H_0=0=H_N$. Since
$$
H_0=H_3=H_2+{\frac12}\Delta f_2=H_0+{\frac12}
\Delta(f_0+f_1+f_2),
$$
we obtain for non-singular doubly periodic 
functions~$f_j$:
$$
f_0+f_1+f_2=C.
$$
Since
$$
e^{f_0}=e^{f_3}=e^{f_2}+H_3=e^{f_2}+H_0+{\frac12}
\Delta(f_0+f_1+f_2)=e^{f_2}+H_0,
$$
we come to the conclusion: if $H_0=0$, then we have
$$
e^{f_3}=e^{f_2}=e^{f_0}.
$$
Hence it follows that
$$
\gather
2f_0+f_1=C,\qquad f_1=C_2f_0,
\\
e^{f_2}=e^{f_1}+H_0+{\frac12}\Delta(f_0+f_1)=
e^{f_0}=e^{f_1}+{\frac12}\Delta(f_0+f_1).
\endgather
$$

Finally, we have
$$
\gather
e^{f_0}-e^{f_1}={\frac12}\Delta(f_0+f_1),
\\
e^{f_0}-e^{C-2f_0}={\frac12}\Delta(f_0+C-2f_0)=
-{\frac12}\Delta f_0.
\endgather
$$

Let $f_0=g+a$. We obtain
$$
e^ae^g-e^{C-2g-2a}=-{\frac12}\Delta g.
$$
If $a=C-2a$, that is, $a=C/3$, then we finally have
$$
e^{C/3}(e^g-e^{-2g})=-{\frac12}\Delta g,
$$
or, after stretching the scales,
$$
\Delta g=e^{-2g}-e^g.
$$
\endexample

The following result was proved in~\cite{30},~\cite{31}.

\proclaim{Theorem~2}
Suppose we are given a cyclic chain of Laplace 
transformations
$$
\dots,L_0,L_1,L_2,\dots,L_N=L_0
$$
such that the operators of the chain have smooth 
doubly periodic real magnetic field and potential. 
Then all the operators of the chain are topologically 
trivial and algebro-geometric with respect to the 
zero level\rom: $L_j\psi=0$. The Bloch functions of 
zero level with all complex quasimomenta are defined 
and sweep out a Riemann surface of finite genus\rom; 
the coefficients of the operators $L_j$ are expressed 
in terms of theta-functions of this surface 
\rom(the `Fermi-curve'\rom), which is common for 
all operators of the chain.
\endproclaim

We recall that two-dimensional algebro-geometric 
operators were introduced and studied 
in~\cite{32},~\cite{33}; for a review of their theory 
see~\cite{34}.

The idea of the proof of this theorem can be seen 
for $N=2$, where the requirement of cyclicity of the 
chain under the conditions of the theorem reduces 
to the equation $\Delta g=\sinh g$.

In studying surfaces with the topology of a torus 
in $\mathbb R^3$ with constant mean curvature it 
was established that all doubly periodic non-singular 
solutions of this equation are finite-zone 
(algebro-geometric); see the survey~\cite{35}. 
Finally, this follows from the fact that any manifold 
of solutions of a non-linear elliptic equation on a 
torus (compact manifold) has finite dimension at a 
non-singular point. Further arguments, which lead to 
the finite-zone property of solutions of the equation 
$\Delta g=\sinh g$, follow the standard 
scheme that was already developed in 1974 for the KdV 
equation (see the survey~\cite{2}); commutative fluxes 
are linearly dependent on a finite-dimensional invariant 
submanifold.

A natural generalization of this idea for all $N\ge2$ 
also leads to the proof of the theorem, as was noted by 
the authors of~\cite{30},~\cite{31}.

\definition{Definition~7}
A {\it semicyclic Laplace chain} is a set of 
Schr\"odinger operators
$$
L_0,\dots,L_N,\qquad L_{j+1}=\widetilde L_j
$$
such that $L_N=L_0+C$.
\enddefinition

\definition{Definition~8}
A {\it quasicyclic Laplace chain} is a set of 
Schr\"odinger operators
$$
L_0,\dots,L_N,\qquad L_{j+1}=\widetilde L_j,
$$
such that the extreme operators are completely 
factorized with a constant potential:
$$
\gather
2L_0=Q_{01}Q_{02}+2V_0,\qquad 2L_N=Q_{N1}Q_{N2}+2V_N,
\\
V_0=\const,\qquad V_N=\const,
\\
Q_{j2}=\overline\partial+B_j(z,\overline z),\qquad
Q_{j1}=\partial+A_j(z,\overline z).
\endgather
$$
>From the definition we have
$$
\gather
2L_j=Q_{j2}Q_{j1}+2W_j,\qquad W_0=-H_0,\qquad W_N=-H_N,
\\
2\widetilde L_j=W_jQ_{j1}^+W_j^{-1}Q_{j2}+2W_j.
\endgather
$$
\enddefinition

\example{Example~5}
$N=1$. A semicyclic chain of length~1. We have
$$
\gather
2L_1=2L_0-2C,\qquad H_1=H_0,
\\
e^{f_1}=e^{f_0}+C=e^{f_0}+H_1,\qquad H_1=C.
\endgather
$$
Hence we obtain
$$
\Delta f_0=0.
$$
In the non-singular doubly periodic case it follows that
$$
f_0=\const,\qquad H=C.
$$
Hence we arrive at a Landau operator in a homogeneous 
constant magnetic field $H=C$ and zero potential. 
Starting from the solutions (let~$C>0$)
$$
Q_{02}\psi_0=0,\qquad L_0\psi_0=0,\qquad V_0=0,
$$
we obtain all the Landau levels
$$
\gather
(Q_{01})\psi_0=\psi_1,
\\
2L_1\psi_1=0\ \Longrightarrow\ 2L_0\psi_1=2C\psi_1,
\enskip C=H.
\endgather
$$
Hence, from the lowest level $Q_0\psi_0=0$ we obtain 
the first level $\psi_1=Q^+_0\psi_0$. These spaces 
of solutions are infinite-dimensional on every level, 
called `Landau levels' for the Landau operator
$$
H+2L_0=Q_{02}Q_{01}+H,\qquad Q_{02}\psi_0=0.
\tag 78
$$
Its levels are
$$
\lambda_n=H+2nH,\quad\psi_n=(Q_{01})^n\psi_0.
\tag 79
$$
\endexample

\example{Example~6}
$N=2$. A semicyclic chain of length~2. Here we have
$
H_2=H_0
$.
In the non-singular doubly periodic case, from the 
equality
$$
H_2=H_0+{\frac12}\Delta(f_0+f_1)
$$
it follows that
$$
\Delta(f_0+f_1)=0\ \Longrightarrow\ f_0+f_1=a.
$$
Since
$$
\gather
e^{f_2}=e^{f_0}+C=e^{f_1}+H_2=e^{f_1}+H_0,
\\
e^{f_1}=e^{f_0}+H_0+{\frac12}\Delta f_0,
\endgather
$$
we obtain
$$
\gather
e^{f_0}-e^{f_1}=H_0-C,
\\
-e^{f_1}+e^{f_0}=-H_0-{\frac12}\Delta f_0.
\endgather
$$

Finally, we have
$$
\gather
2(e^{f_0}-e^{f_1})=-C-{\frac12}\Delta f_0,
\\
f_1=a-f_0.
\endgather
$$
Thus we arrive at the equation
$$
e^{f_0}-e^{a-f_0}+\frac{C}{2}=-\frac{1}{4}\Delta f_0.
$$

Let $f_0=g+a/2$. We have the equality
$$
\gather
e^{g+a/2}-e^{a/2-g}+\frac{C}{2}=-\frac{1}{4}\Delta g,
\\
-2C-2e^{a/2}\sinh g=\Delta g.
\endgather
$$
This equation has an extensive family of doubly 
periodic non-singular real solutions.
\endexample

Solutions of the equation $L_N\psi_N=(L_0+C)\psi_N=0$ 
are obtained from solutions of the equation 
$L_0\psi_0=0$ by the chain of algebraic `creation 
operators'
$$
\psi_N=Q_{N-1,2}\dotsb Q_{0,2}(\psi_0),
\qquad L_0\psi_N=(-C)\psi_N.
$$
Hence, the levels of spectrum $\varepsilon=0$ 
and $\varepsilon=-C$ for the operator $L_0$ are 
connected by these operators and are obtained from 
each other. It is easy to see that magnetic Bloch 
solutions are transformed into magnetic Bloch ones 
under the action of these operators if the magnetic 
flux is quantized (that is, the Bloch solutions are 
defined).

\proclaim{Lemma~9}
If $C\ne0$, then the magnetic flux of a non-singular 
semicyclic chain with potentials that do not vanish 
is different from zero.
\endproclaim

\remark{Remark~$6$}
Since $H_{j+1}=H_j+{\frac12}\Delta(f_j)$ and $f_j$ 
is non-singular, we see that the fluxes of the 
operators $L_j$ coincide: $[H_{j+1}]=[H_j]$.
\endremark

\demo{Proof of the lemma}
Since
$$
e^{f_N}=e^{f_{N-1}}+H_0+{\frac12}\Delta
\biggl(\sum_{j=0}^{N-1}f_j\biggr)
=e^{f_0}+C,
$$
and $[e^{f_N}]=[e^{f_{N-1}}]+[H]$, we obtain
$$
[e^{f_N}]=[e^{f_0}]+[C]=[e^{f_0}]+KC=[e^{f_0}]+N[H].
$$
Hence, for the flux $N[H]=N[H_0]=KC$ it follows 
that
$$
C=\frac{N[H]}{K}\,,
\tag 80
$$
where $K$ is the area of an elementary cell. We 
have proved Lemma~9.
\enddemo

(For purely cyclic non-singular chains we have $C=0$, 
that is, the flux is always equal to zero.)

The greatest interest is in quasicyclic chains, with 
which the main results of \cite{30},~\cite{31} are 
connected.

\proclaim{Theorem~3}
Let the quasicyclic Laplace chain of the Schr\"odinger 
operators
$$
\gather
L_0,\dots,L_N,\qquad L_{j+1}=\widetilde L_j,
\\
2L_0=Q_{0,1}^+Q_{0,2},\qquad 2L_N=Q^+_{N,1}Q_{N,2}+2C_N
\endgather
$$
be given with doubly periodic real quantities 
$(f_j,H_j)$, and suppose that the magnetic flux 
$[H_j]$ is positive, $[H]>0$. Then the operator 
$L_N$ has two strongly degenerate discrete levels, 
which are isomorphic to the Landau levels and one of 
them is the lowest\rom:
\roster
\Item"\rom{a)}" $L_N\psi=C_N\psi$, $Q_N\psi=0$, 
$\lambda=C_N$,
\Item"\rom{b)}" $L_N\psi=0$, $\lambda=0$.
\endroster
In addition, $C_N=N[H]/K$, where $K$ is the area of 
an elementary cell.
\endproclaim

Here $Q_{j2}=\overline\partial+B_j(z,\overline z)$,
$Q_{j1}=\partial+A_j(z,\overline z)$.

\demo{Proof}
First, solving the equation
$$
Q_{N,2}\psi=0
$$
according to the scheme of~\cite{31}, we obtain the 
eigenvalue $L_N\psi=C_N\psi$. If $[H]>0$, 
such a solution exists and gives a huge space 
isomorphic to the Landau level according to~\cite{7}, 
\cite{8},~\cite{36}. A~basis in this space, mentioned 
in~\cite{7}, \cite{8},~\cite{36}, is the magnetic 
Bloch basis, which is suitable only in the case of 
the integer-valued flux $[H]=2\pi m>0$; however, 
the result itself is true for all positive values 
of the flux.

We obtain the second level as follows. We solve the 
equation $Q_{02}\psi=0$ according to~\cite{7},~\cite{8} 
and obtain also for $[H]>0$ a huge space of functions 
$\psi$ isomorphic to the Landau level, which can be 
defined by using elliptic formulae for the integer-valued 
flux~$[H]$, common for $H_0$ and~$H_N$,
$$
[H]=[H_0]=[H_N].
$$
Next we factorize the operator $L_0$ in the form
$$
2L_0=Q_{0,2}Q_{0,1}+2H_0=Q_{0,1}^+Q_{0,2}.
$$
Then we apply the creation operator $N$ times to a 
solution of $Q_0\psi=0$:
$$
\widehat\psi=Q_{N-1,1}\circ\dots\circ Q_{0,1}\psi.
$$
The new function $\widehat\psi$ satisfies the equation
$$
L_N\widehat\psi=0=(Q_{N,1}Q_{N,2}+C_N)\widehat\psi.
$$
The function $\widehat\psi$ belongs to $\script L_2
(\mathbb R)$ if $\psi$ belongs to~$\script L_2(\mathbb 
R)$.

Now we find the number~$C_N$. We have for the Laplace 
chain the following boundary conditions:
$$
-e^{f_0}=H_0,\qquad -e^{f_N}=H_N+C_N.
$$
For the mean values we obtain
$$
[e^{f_{j+1}}]=[e^{f_j}]+[H_{j+1}]=[e^{f_j}]+[H]=
[e^{f_0}]+(j+1)[H]=(j+2)[H].
$$
In particular, for $j+1=N$ we obtain
$$
e^{f_N}=[H]+[C_N]=(N+1)[H],
$$
where $[C_N]=KC_N$ or $C_N=N[H]/K$.
\enddemo

\remark{Remark~$7$}
Let the operator $L$ be given in the self-adjoint form,
$$
\gather
Q=Q_1=\partial+A,\qquad Q_2=-Q^+=\overline\partial+B,
\\
\overline B=-A.
\endgather
$$
It is also convenient to choose the `Lorentz gauge', 
for which all the formulae are written 
in~\cite{7},~\cite{8}:
$$
\Im(A_{\overline z})=0.
\tag 81
$$
Then we have
$$
2L=-Q^+Q+2W=-QQ^++2V.
$$
\endremark

Further, if $L\psi=0$, we carry out the Laplace 
transformation together with the gauge transformation
$$
\gather
L\longmapsto e^g\widetilde Le^{-g},\qquad W=e^f,
\\
\widetilde L=-WQW^{-1}Q^++2W,
\\
\widetilde\psi\longmapsto e^gQ\psi=e^g\widetilde\psi.
\endgather
$$
Choosing $g=-f/2$, we obtain the operator 
$\widetilde{\widetilde L}=e^{-f/2}\widetilde Le^{f/2}$ 
in the form
$$
\widetilde{\widetilde L}=e^{-f/2}\widetilde Le^{f/2}
=-\widetilde Q\widetilde Q^++2W,
$$
where $\widetilde Q=e^{-f/2}Qe^{f/2}$, $\widetilde 
Q^+=e^{f/2}Q^+e^{-f/2}$. We obtain explicitly the 
self-adjoint operator in the Lorentz gauge 
transformation.

\example{Example~7}
Let $N=1$. From the definition for $L_0$ we have
$$
2L_0=-Q_0^+Q_0=-Q_0Q_0^++2H_0,
$$
where
$$
Q_0=\partial+A,\qquad -Q_0^+-\overline\partial+B,\qquad
H_0=W_0=e^{f_0}.
$$
Therefore
$$
2L_1=-(e^{f_0/2}Q_0^+e^{f_0/2})(e^{f_0/2}Q_0
e^{-f_0/2})+2H_0=
\widetilde{\widetilde L_0}=e^{-f_0/2}\widetilde 
L_0e^{f_0/2}.
$$
>From the quasicyclicity condition for $N=1$ we have
$$
H_0=\const,\qquad L_1=L_0+2H_0.
$$
We arrive at the Landau operator.
\endexample

\example{Example~8}
Let $N=2$.
$$
\gather
2L_0=-Q_0^+Q_0=-Q_0Q_0^++2H_0,\qquad H_0=e^{f_0},
\\
2L_2=-Q_2^+Q_2+2C_2=-Q_2Q_2^++2C_2+2H_2,
\\
e^{f_2}=C_2+H_2=e^{f_1}+H_2.
\endgather
$$
We see that $e^{f_1}=\const$. Further,
$$
C_2=e^{f_1}=e^{f_0}+H_1=e^{f_0}+H_0+{\frac12}\Delta f_0,
$$
where $H_0=e^{f_0}$. Hence we have
$$
\gather
C_2-2e^{f_0}={\frac12}\Delta f_0,
\\
2C_2=-4e^{f_0}=\Delta f_0.
\endgather
$$
Putting $f_0=g+a$, where $e^a=C_2/2$, we obtain 
$\Delta g=e^a(1-e^g)$. By stretching the axes we 
bring this equation to the form
$$
\Delta g=1-e^g.
\tag 82
$$
The last equation has an extensive family of smooth 
real doubly periodic solutions. Even its solutions, 
dependent on one variable (for example, on~$x$), 
which can be found by quadrature, give us a continuous 
family belonging to this class, with an arbitrary 
period in~$y$:
$$
\gather
f_0''=2C_2-4e^{f_0},\qquad C_2>0,
\\
\tau''=-\frac{\partial}{\partial\tau}(U(\tau)),
\\
U(\tau)=-A\tau+4e^\tau,\qquad\tau=f_0,\qquad A=2C_2.
\endgather
$$
The dependence of $f_0(x)$ is determined by quadrature:
$$
\gather
x-x_0=\int\frac{d\tau}{\sqrt{-4e^\tau+A\tau+E}}\,,
\\
E\ge E_{\min}=A\Bigl(1-\log\frac{A}{4}\Bigr).
\endgather
$$

Equation (82) has occurred in various physical 
problems---in plasma theory, superconductivity 
theory, and so on. Hector~de~Vega informed one of 
the authors (S.~P.~Novikov) that equation (82) 
arose as the Bogomol'nyi reduction for the 
Ginzburg--Landau equation for the critical value 
of the parameter which separates superconductors 
of types~I and~II (see~\cite{37}). However, the 
application of this form of the Ginzburg--Landau 
equation at a critical point is not clear.
In addition, the magnetic field in this model 
differs from ours by a non-zero constant, and 
in the theory of superconductors solutions with 
singularities are necessary.
\endexample

\remark{Hypothesis}
For periods $N\ge5$ the quasiperiodicity condition 
does not give smooth real doubly periodic solutions, 
apart from a constant. In any case, the constant (the 
Landau operator) does not allow any variation---it 
is an isolated solution.
\endremark

According to the hypothesis of Novikov, the 
resulting class of operators is `maximally exactly 
soluble' among two-dimensional Schr\"odinger operators 
with smooth real doubly periodic electric potential 
and magnetic field with a non-zero (let it be 
integer-valued) flux~$[H]=2\pi m$.

These operators have two discrete levels with 
infinite degeneration that are isomorphic to the 
Landau levels. One of those levels is basic (that 
is, the lowest one) and the second one is certainly 
not the nearest to the first one (at least one 
`magnetic zone' lies between them). If there are 
three strongly degenerate discrete levels, then, 
as we suppose, the operator will coincide with the 
Landau operator.

The general arrangement of the spectrum of 
Schr\"odinger operators in a magnetic field with 
integer-valued flux, interesting aspects of the 
inverse spectral problem, and geometric and topological 
properties of the spectrum of these operators were 
discussed in~\cite{15}, \cite{38},~\cite{39}.

If the magnetic flux is quantized, then we have the 
`magnetic Bloch function' of the dispersion law with 
number $m\ge0$, $m\in\mathbb Z$
$$
\widehat T_1\psi=e^{ip_1T_1}\psi,\qquad
\widehat T_2\psi=e^{ip_2T_2}\psi,\qquad
L\psi_m=\varepsilon_m(p_1,p_2)\psi_m,
\tag 83
$$
$\psi=\psi_m(x,y,p_1,p_2)$, $[H]=\Phi\in2\pi\mathbb Z$, 
$\Phi\ne0$.

For definiteness, let the function $\psi$ be 
normalized by analogy with finite-zone theory~\cite{2}:
$$
\psi_m(x,y,p_1,p_2)\bigr|_{x=x_0,y=y_0}=1.
$$
The function $\psi_m(x,y,x_0,y_0,p_1,p_2)$ under 
the fixed normalization $(x_0,y_0)$ is defined on 
the `phase space'
$$
(x,y,p_1,p_2)\in\mathbb R^2\times\mathbb T^2.
\tag 84
$$
Moreover, the function $\psi$ is defined up to 
normalization
$$
\psi_n\mapsto f(p_1,p_2)\psi_n.
\tag 85
$$
The manifold of zeros of the function $\psi_m$, 
$\psi_m=0$, splits into two parts:

1) `topological zeros', where $\psi_m(x,y)\equiv0$ 
for fixed $p_1$,~$p_2$; this part of the zeros 
changes under transformations~(85), but the 
`algebraic number of zeros' is the `Chern class' 
$c_1^{(m)}$ of the dispersion law with number $m$ 
for the operator $L$ of general position, where the 
dispersion laws do not intersect,
$$
\varepsilon_{m_1}(p_1,p_2)\ne\varepsilon_{m_2}
(p_1,p_2),\qquad
m_1\ne m_2;
$$

2) `geometric zeros', where locally at a general 
point we can assume that
$$
(p_1,p_2)=\gamma_m(x,y),\qquad m\ge0.
$$
As $m\to\infty$ the dispersion laws converge to constants:
$$
\varepsilon_m(p_1,p_2)\to\const=(2m+1)[H],
$$
analogous to the forbidden zones of the one-dimensional 
periodic Schr\"odinger \linebreak operator.

\remark{Problem}
Find the analogue of the `Dubrovin equations' for 
the set $\{\gamma_m(x,y)\}$ for all~$m$; solve the 
inverse problem of reconstructing the operator~$L$, 
starting from the set of geometric data, in the 
phase space~(84), analogous to the one-dimensional 
data, without entering into the complex domain of 
quasimomenta.
\endremark

\head
\boldsymbol\S5. Difference equations and Laplace 
transformations. The hyperbolic case
\endhead

The simplest hyperbolic difference equation on 
the standard quadratic lattice $\Gamma$ in 
$\mathbb R^2$ is
$$
a_n\psi_n+b_n\psi_{n+T_1}+c_n\psi_{n+T_2}+d_n
\psi_{n+T_1+T_2}=0,
\tag 86
$$
where $n=(n_1,n_2)$ is the number of a lattice point, 
$n_j\in\mathbb Z$, $T_1=(1,0)$, $T_2=(0,1)$ are the 
basis vectors of the lattice. We also denote by 
$T_1$,~$T_2$ the basis operators of translations 
which act on functions defined on~$\mathbb Z^2$:
$$
\gather
T_1\psi(n_1,n_2)=\psi(n_1+1,n_2),
\\
T_2\psi(n_1,n_2)=\psi(n_1,n_2+1).
\endgather
$$
We can write (86) in the following way:
$$
L\psi=(a_n+b_nT_1+c_nT_2+d_nT_1T_2)\psi=0.
\tag 87
$$

It was already observed in~\cite{40} that for an 
equation of this form there is a class of 
algebro-geometric exactly soluble cases with 
effective theta-function formulae.

The operator $L$ is defined up to multiplication 
by a function: $L\mapsto f_nL$. We also 
have gauge transformations, and so together we 
have a system of the following equivalences of~(87):
$$
L\mapsto f_nLg_n,\qquad \psi_n\mapsto g_n^{-1}\psi_n,
\tag 88
$$
where $f_n$, $g_n$ are non-vanishing functions on 
the lattice.

\proclaim{Lemma~10}
The operator $L$ admits a unique representation 
in the form
$$
\text{\rom{a)}}\quad L=f_n\bigl((1+u_nT_1)
(1+v_nT_2)+w_n\bigr)
\tag 89
$$
or in the form
$$
\text{\rom{b)}}\quad L=f_n'\bigl((1+v_n'T_2)
(1+u_n'T_1)+w_n'\bigr).
\tag 90
$$
The coefficients $u_n$, $v_n$, $w_n$, $f_n$ for the 
factorization~\rom{a)} are determined from the formulae
$$
a_n=f_n(1+w_n),\qquad f_nu_n=b_n,\qquad f_nv_n=c_n,\qquad
f_nu_nv_{n+T_1}=d_n.
\tag 91
$$
\endproclaim

The proof is based on direct verification.

Using both factorizations a) and~b), we define two 
Laplace transformations of (87)~$L\psi=0$.

\definition{Definition~9}
A {\it Laplace transformation of the first type} is 
a transformation
$$
\aligned
L&\longmapsto\widetilde L=\widetilde f_n\bigl(w_n
(1+v_nT_2w_n^{-1})
(1+u_nT_1)+w_n\bigr),
\\
\psi&\longmapsto\widetilde\psi=(1+v_nT_2)\psi.
\endaligned
\tag 92
$$
A {\it Laplace transformation of the second type} 
is a transformation
$$
\aligned
L&\longmapsto\widetilde L'=\widetilde f'\bigl(w_n'
(1+u_n'T_1)
(w_n')^{-1}(1+v_n'T_2)+w_n'\bigr),
\\
\psi&\longmapsto\widetilde\psi'=(1+u_n'T_1)\psi.
\endaligned
\tag 93
$$
\enddefinition

\proclaim{Lemma~11}
The Laplace transformations of the first and second 
types are mutually inverse.
\endproclaim

The proof is obtained by simple substitution, as 
in the continuous case.

We denote a Laplace transformation of the first 
type by $\Lambda_{12}^{++}$ and of the second 
type by~$\Lambda_{21}^{++}$. Lemma~11 claims that 
$\Lambda_{12}^{++}\Lambda_{21}^{++}=1$.

It is appropriate to note that in a completely 
analogous way we can define the Laplace 
transformation of~(86), which corresponds to 
an arbitrary pair of ortho\-normal basis translations
$$
\begin{aligned}
(T_1,T_2),\ (T_2,T_1)&\:\ \Lambda_{12}^{++},
\Lambda_{21}^{++},
\\
(T_1^{-1},T_2),\ (T_2,T_1^{-1})&\:\ \Lambda_{12}^{-+},
\Lambda_{21}^{+-},
\\
(T_1,T_2^{-1}),\ (T_2^{-1},T_1)&\:\ \Lambda_{12}^{+-},
\Lambda_{21}^{-+},
\\
(T_1^{-1},T_2^{-1}),\ (T_2^{-1},T_1^{-1})&\:\
\Lambda_{12}^{--},\Lambda_{21}^{--}.
\end{aligned}
$$
Thus, we have eight Laplace transformations 
$\Lambda_{12}^{\varepsilon\sigma}$, 
$\Lambda_{21}^{\varepsilon\sigma}$,
$\varepsilon,\sigma=\pm$, and
$$
\Lambda_{12}^{\varepsilon\sigma}
\Lambda_{21}^{\sigma\varepsilon}=1.
$$
Hence, the Laplace transformations here form a 
group with four generating elements 
$\Lambda_{12}^{\varepsilon\sigma}$, the structure 
of which we do not know so far.

Now we return to the transformation $\Lambda_{12}^{++}$, 
described above. The Laplace transformations are 
well defined on the equivalence classes~(88). 
Therefore, as in the continuous case, they should 
be registered on the invariants of the equivalence 
classes, which are analogous to the magnetic field 
and electric potential.

\proclaim{Lemma~12}
The complete set of invariants with respect to 
the equivalence $(88)$ is given by the following 
basis functions\rom:
$$
K_{1n}=\frac{b_nc_{n+T_1}}{d_na_{n+T_1}}\,,\qquad
K_{2n}=\frac{c_nb_{n+T_2}}{d_na_{n+T_2}}\,.
\tag 94
$$
All other invariants are expressed in terms of 
$K_{1n}$, $K_{2n}$ and their translations on the lattice.
\endproclaim

\demo{Proof}
The proof of the invariance of these quantities 
is straightforward: it is a direct elementary 
verification. It is convenient to carry out the 
proof of their completeness for the factorized 
form~(89). Expressing $a_n,b_n,c_n,d_n$ in terms 
of $f_n,u_n,v_n,w_n$, we obtain
$$
K_{1n}=\frac{1}{1+w_n}\,,\qquad K_{2n}=\frac{1}{1+w_{n+T_2}}
\bigl(v_nu_{n+T_2}v_{n+T_1}^{-1}u_n^{-1}\bigr).
$$
It is clear that the potential $w_n$ is gauge 
invariant, and the function $f_n$ can be shortened 
in equivalence classes. Hence, everything reduces 
to gauge equivalence for $f_n=1$. Here the 
`potential' $w_n$ and the `curvature'
$$
H=v_nu_n^{-1}u_{n+T_2}v_{n+T_1}^{-1}=\frac{v_nu_{n+T_2}}{u_nv_{n+T_1}}
\tag 95
$$
are gauge invariant. We need to find all gauge 
invariants of the operator
$$
L_0=(1+u_nT_1)(1+v_nT_2)
$$
with respect to the transformations
$$
L_0\longmapsto\tau_n(1+u_nT_1)(1+v_nT_2)\tau_n^{-1}=
\biggl(1+u_n\frac{\tau_n}{\tau_{n+T_1}}T_1\biggr)
\biggl(1+v_n\frac{\tau_n}{\tau_{n+T_2}}T_2\biggr).
$$
This problem is easily reduced to the classification 
of gauge invariant monomials, that is, products of 
the quantities $u_{n+kT_1+lT_2}$, $v_{n+kT_1+lT_2}$ 
for different (finitely many)~$k$,~$l$. Elementary 
combinatorics leads to the conclusion that all these 
monomials are products of monomials (94) and their 
translations in some powers. We omit these details.
\enddemo

We use the basis of invariants $w_n$, $H_n$, by 
analogy with the continuous case.

\proclaim{Lemma~13}
The Laplace transformation for invariants has the form
$$
\gathered
1+\widetilde w_{n+T_1}=(1+w_{n+T_2})
\frac{w_nw_{n+T_1+T_2}}{w_{n+T_1}w_{n+T_2}}H_n^{-1},
\\
\widetilde H_n=\frac{1+w_{n+T_2}}{1+\widetilde 
w_{n+T_2}}\,.
\endgathered
\tag 96
$$
\endproclaim

The proof is obtained by substituting the expressions 
for $\widetilde K_1$, $\widetilde K_2$ in terms 
of \linebreak$\widetilde a$, $\widetilde b$, 
$\widetilde c$,~$\widetilde d$.

We now consider an infinite chain of Laplace 
transformations for the quantities
$$
\widetilde H_n^{(k)}=H_n^{(k+1)},\qquad
\widetilde w_n^{(k)}=w_n^{(k+1)},
$$
where $n=(n_1,n_2)$, $k\in\mathbb Z$, and 
$H_n^{(k+1)}$,~$w_n^{(k+1)}$ are obtained from 
the previous ones by a Laplace transformation. 
Eliminating $H_n^{(k)}$ from the equations, we 
obtain a `completely discrete analogue of the 
two-dimensionalized Toda chain':
$$
\frac{1+w_{n+T_1}^{(k+2)}}{1+w_{n+T_1}^{(k+1)}}\,
\frac{1+w_{n+T_2}^{(k+1)}}{1+w_{n+T_2}^{(k)}}=
\frac{w^{(k)}_{n+T_1}w^{(k)}_{n+T_2}}{w^{(k)}_n
w^{(k)}_{n+T_1+T_2}}\,.
$$

\remark{Problem}
Find a comparison between this completely discrete 
analogue of the two-dimensionalized Toda chain and 
the systems studied in~\cite{41}, starting from the 
theory of the `Yang--Baxter' equation.
\endremark

\example{Example~9}
We consider a cyclic chain of length $N=2$. After 
simple calculations we obtain
$$
\gather
w_n^{(1)}=C(w_n^{(0)})^{-1},
\\
w_{n+T_1+T_2}=w_n^{-1}(C+w_{n+T_1})(C+w_{n+T_2})
(1+w_{n+T_1})^{-1}
(1+w_{n+T_2})^{-1}.
\endgather
$$
\endexample

Finally, we note that there is, as we showed above, 
the group of all hyperbolic Laplace transformations, 
generated by four basis transformations 
$\Lambda_{12}^{\varepsilon\sigma}$, 
$\varepsilon,\sigma=\pm$, which corresponds to 
the choice of the bases $(T_1^\varepsilon,T_2^\sigma)$. 
In principle, to each word in this group corresponds 
the condition for cyclicity of a chain. However, 
the structure of this group is not known so far.

\head
\boldsymbol\S6. The Laplace transformation for 
elliptic two-dimensional difference operators on 
a regular lattice. Equations of a triangle, curvature
\endhead

It appears that in the elliptic case Laplace 
transformations cannot be defined for difference 
operators of the second order on a square lattice. 
To give their correct definition we should follow 
two principles.

I. In order that continuous and discrete 
`spectral symmetries' of continuous Schr\"odinger 
operators be conserved in the difference case, all
translations should be considered as covariant, as 
in the one-dimensional case (see~\S3).

II. In the two-dimensional case for elliptic 
operators we should replace the square lattice 
by an equilateral triangular lattice, where the 
following translations have the same length:
$$
T_1^{\pm1},\ T_2^{\pm1},\ (T_1T_2)^{\pm1}.
\tag 97
$$
We denote an arbitrary pair of neighbouring 
translations~(97) (which go anticlockwise) by 
$T_1^*$,~$T_2^*$ and ascribe to this pair a 
Laplace transformation generated by factorization
$$
L=(x_n+y_nT_1^*+z_nT_2^*)(x_n+y_{n-T_1^*}
{T_1^*}^{-1}+z_{n-T_2^*}
{T_2^*}^{-1})+w_n,
\tag 98
$$
where the self-adjoint operator $L$ has the form
$$
L=a_n+b_nT_1+c_nT_2+d_{n-T_2}T_1T_2^{-1}+
b_{n-T_1}T_1^{-1}+c_{n-T_2}
T_2^{-1}+d_{n-T_1}T_2T_1^{-1},
\tag 99
$$
all the coefficients are real, and $b_n,c_n,d_n\in
\mathbb R_+$. It is easy to see that the coefficient 
$a_n$ is a numerical function on~$\mathbb Z^2$, 
while the coefficients $b_n,c_n,d_n$ are a connection 
sitting on the corresponding edges of the lattice. 
As a result the operator contains an interaction of 
each vertex with all six nearest neighbours, and all 
edges have the same length. We start from the 
following observation~(\cite{42},~\cite{43}).

\proclaim{Lemma~14}
A real self-adjoint operator of the form $(99)$ 
with non-zero coefficients $b_n$,~$c_n$,~$d_n$ 
always admits a unique representation in the form
$$
L=QQ^++w_n,\qquad Q=x_n+y_nT_1+z_nT_2,
\tag 100
$$
where $T_1^+=T_1^{-1}$, $T_2^+=T_2^{-1}$, $(AB)^+=
B^+A^+$. Moreover we have the equalities
$$
\aligned
x_ny_{n-T_1}&=b_{n-T_1},
\\
x_nz_{n-T_2}&=c_{n-T_2},
\\
y_{n-T_1}z_{n-T_2}&=d_{n-T_1-T_2}.
\endaligned
\tag 101
$$
\endproclaim

In all, following~\cite{42},~\cite{43}, we obtain six 
different factorizations, which correspond to the following 
six pairs of periods $(T_1^*,T_2^*)$: 1)~$(T_1,T_2)$, 
2)~$(T_2,T_1^{-1}T_2)$, 3)~$(T_1^{-1}T_2,T_1^{-1})$, 
4)~$(T_1^{-1},T_2^{-1})$, 5)~$(T_2^{-1},T_2^{-1}T_1)$, 
6)~$(T_2^{-1}T_1,T_1)$,
$$
L=Q_jQ_j^++w_{jn},
$$
$j=1,2,3,4,5,6$. $j$ is determined modulo~6.

\definition{Definition~10}
A {\it Laplace transformation $P_j$ of type $j$} is as follows:
$$
\aligned
&P_j\:L\longmapsto\widetilde L^{(j)}=w_{jn}^{1/2}Q_j^+
w_{jn}^{-1}Q_jw_{jn}^{1/2}+w_{jn},
\\
&P_j\:\psi\longmapsto\widetilde\psi=w_{jn}^{-1/2}Q_j^+\psi.
\endaligned
\tag 102
$$
It is defined if $w_{jn}>0$.
\enddefinition

\proclaim{Lemma~15}
The Laplace transformations $P_j$ and $P_{j+3}$ are 
mutually inverse. The others are connected by the following 
relations\rom:
$$
\gathered
w_{1n}=w_{3n}=w_{5n},
\\
Q_1=Q_3T_1=Q_5T_2,
\\
\widetilde L^{(1)}=\sqrt{\frac{w_n}{w_{n-T_1}}}\,
T_1^{-1}\widetilde L^{(3)}T_1\sqrt{\frac{w_n}{w_{n-T_1}}}
=\sqrt\frac{{w_n}{w_{n-T_2}}}\,
T_2^{-1}\widetilde L^{(5)}T_2\sqrt{\frac{w_n}{w_{n-T_2}}}\,.
\endgathered
\tag 103
$$
In the equivalence class
$$
L\mapsto f_nLf_n
\tag 104
$$
the operators $\widetilde L^{(1)}$, $\widetilde L^{(3)}$, 
$\widetilde L^{(5)}$ are unitarily adjoint.

For $j=2,4,6$ analogous formulae hold\rom:
$$
\widetilde L^{(4)}=\sqrt{\frac{w_n}{w_{n+T_1}}}\,
T_1\widetilde L^{(6)}T_1^{-1}\sqrt{\frac{w_n}{w_{n+T_1}}}
=\sqrt{\frac{w_n}{w_{n+T_2}}}\,T_2\widetilde L^{(2)}T_2^{-1}
\sqrt{\displaystyle\frac{w_n}{w_{n+T_2}}}\,.
$$
\endproclaim

\remark{Remark~$8$}
In the definition of Laplace transformations we can discard 
the requirement $w_n>0$, replacing it by the condition $w_n\ne0$, and set
$$
\widetilde L^{(j)}=Q_j^+w_{jn}^{-1}Q_j+1,\qquad
\widetilde\psi=Q_j^+\psi.
$$
We obtain the same transformations up to equivalence~(104), 
but $P_j$ and $P_{j+3}$ are not exactly mutually inverse. We 
note that in the discrete case, as opposed to the continuous 
one, the condition $w\ne0$ does not mean that $w$ has constant sign.
\endremark

\remark{Remark~$9$}
The relations (103) were not noted in~\cite{42}--\cite{44}.
\endremark

\remark{Remark~$10$}
If the potential $w_n\equiv w_0$ is constant (for example, for 
the first factorization), then the Laplace transformation
$$
P_1\:L\mapsto\widetilde L^{(1)}
$$
acts on the eigenfunctions of all levels $L\psi=\lambda\psi$:
$$
P_1\psi=Q_1^+\psi,\qquad P_1L=Q_1^+Q_1+w_0=\widetilde L^{(1)}.
$$
In the framework of the equivalence class (104) of real 
self-adjoint operators we can always arrange that $w_n=\const$ 
if $w_n=\exp f_n\ne0$. However, these transformations realize 
a formal equivalence of the spectral theories only on the zero 
level $L\psi=0$, and therefore we can apply the 
Laplace transformation to all levels, generally speaking, only 
once (for special cases see below).
\endremark

The proof of the lemma can be obtained by direct verification.

An investigation of the cyclic chains of difference operators 
has not been carried out. The classes of purely factorized 
operators of the form~a) and~b):
$$
\text{a)}\ \ L=Q^+Q;\qquad
\text{b)}\ \ \widetilde L=QQ^+
$$
are of interest, where
$$
Q=1+ce^{l_1(n)}T_1+de^{l_2(n)}T_2,
\tag 105
$$
and
$$
l_j(n)=l_{j1}n_1+l_{j2}n_2
$$
are linear forms in $n=(n_1,n_2)$.

Following~\cite{43},~\cite{44}, we consider the `equation 
of black triangles', which have the form $\langle n,n+T_1,n+T_2\rangle$:
$$
Q\psi=0,
$$
and the `equation of white triangles', which have the form 
$\langle n,n-T_1,n-T_2\rangle$:
$$
Q^+\psi=0.
$$

\proclaim{Theorem~4}
$1)$ The equation of black triangles $Q\psi=0$ 
certainly has an infinite-dimensional space of solutions 
$\psi\in\script L_2(\mathbb Z^2)$ if one of the following 
conditions is satisfied\rom:
\roster
\Item"\rom{a)}" $l_{11},l_{22}>0$, 
$l_{11}l_{22}-l_{12}^2>0;$
\Item"\rom{b)}" $l_{11},l_{22}>0$, 
$l_{11}l_{22}-l_{21}^2>0;$
\Item"\rom{c$'$)}" $l_{11}>0$, 
$l_{11}l_{22}-l_{12}^2>l_{11}(l_{21}-l_{12});$
\Item"\rom{c}$'')$" $l_{22}>0$, 
$l_{11}l_{22}-l_{21}^2>l_{22}(l_{21}-l_{12})$.
\endroster

$2)$ The equation of white triangles $Q^+\psi=0$ 
certainly has an infinite-dimensional space of solutions 
$\psi\in\script L_2(\mathbb Z^2)$ if one of the following 
conditions, which are obtained from the previous ones by 
the transformation $l_{ij}\to-l_{ij}$, is satisfied\rom:
\roster
\Item"\rom{a)}" $l_{11},l_{22}<0$, $l_{11}l_{22}-l_{12}^2>0;$
\Item"\rom{b)}" $l_{11},l_{22}<0$, $l_{11}l_{22}-l_{21}^2>0;$
\Item"\rom{c$'$)}" $l_{11}<0$, 
$l_{11}l_{22}-l_{12}^2>l_{11}(l_{12}-l_{21});$
\Item"\rom{c}$'')$" $l_{22}<0$, 
$l_{11}l_{22}-l_{21}^2>l_{22}(l_{12}-l_{21})$.
\endroster
\endproclaim

\demo{Proof}
We construct explicitly the solutions of our equations 
if the conditions presented above are satisfied. We examine 
first the case of black triangles. We are looking for a 
solution of the form
$$
\psi_n=e^{-K_2(n)}\chi_n,
$$
where
$$
K_2(n)=\alpha n_1^2+2\beta n_1n_2+\delta n_2^2.
$$
For $\chi_n$ we obtain the equation
$$
0=\chi_n+ce^{-\alpha}e^{(l_{11}-2\alpha)n_1+
(l_{12}-2\beta)n_2}\chi_{n+T_1}+
de^{-\delta}e^{(l_{21}-2\beta)n_1+(l_{22}-2\delta)n_2}\chi_{n+T_2}.
\tag 106
$$
We consider three cases:
\roster
\Item"a)" the coefficients (106) depend only on~$n_1$;
\Item"b)" the coefficients (106) depend only on~$n_2$;
\Item"c)" the coefficients (106) depend only on~$n_1+n_2$.
\endroster
This leads, respectively, to the conditions:
\roster
\Item"a)" $l_{12}=2\beta$, $l_{22}=2\delta$;
\Item"b)" $l_{11}=2\alpha$, $l_{21}=2\beta$;
\Item"c)" $l_{11}-2\alpha=l_{12}-2\beta$, $l_{21}-2\beta=l_{22}-2\delta$.
\endroster
In the case~a) we make the substitution
$$
\chi_n=w^{n_2}\varphi_{n_1}
\tag 107
$$
and obtain the equation for~$\varphi_m$:
$$
0=\varphi_m+ce^{-\alpha}e^{(l_{11}-2\alpha)(m+1)}\varphi_{m+1}+
wde^{-\delta}e^{(l_{21}-l_{12})m}\varphi_m.
$$
In addition we put $l_{11}=2\alpha$. After this we obtain
$$
(1+wce^{-l_{22}/2}e^{(l_{21}-l_{12})m})\varphi_m=-ce^{-\alpha}
\varphi_{m+1}.
\tag 108
$$
Two cases are possible.

\smallskip
\noindent{\sl Case}~1. $l_{21}>l_{12}$. We choose the value 
of $w_q$ such that
$$
1+w_qde^{-\delta/2}e^{(l_{21}-2\beta)q}=0,
\tag 109
$$
where $2\delta=l_{22}$, $2\beta=l_{12}$. 
>From this we obtain the `quantization condition', 
which selects a discrete series of admissible values $w=w_q$, 
where $q$ runs over~$\mathbb Z$. Under this condition 
$\varphi_m^{(q)}=0$ for~$m>q$.

Further, we consider the relation
$$
\varphi_m^{(q)}=-(ce^{-\alpha})(1+w_q
e^{k_2-\delta}e^{(l_{21}-2\beta)m})^{-1}
\varphi_{m+1}
$$
as $m\to\infty$. If $l_{21}-2\beta>0$, then for $m<0$ and 
$|m|\to\infty$ we have the asymptotics
$$
\varphi_m^{(q)}\sim(\const)^{|m|}.
$$
Our solution is constructed in the form
$$
\psi_n^{(q)}=e^{-K_2(n)}w^{n_2}\varphi_{n_1}^{(q)},
\tag 110
$$
where $2K_2(n)=l_{11}n_1^2+2l_{12}n_1n_2+l_{22}n_2^2$. 
This form is strictly positive if and only if
$$
l_{11}>0,\qquad l_{22}>0,\qquad l_{11}l_{22}-l_{12}^2>0.
$$
The function $\psi_n^{(q)}$ certainly belongs to the space 
$\script L_2(\mathbb Z^2)$ and takes the value zero for~$n_1>q$.

\smallskip
\noindent{\sl Case}~2. $l_{12}\ge l_{21}$. In this case a 
solution $\varphi_m$ of equation (108) grows no faster 
than $(\const)^{|m|}$, $m\to\pm\infty$, for any~$w$. 
Thus we have constructed the necessary solution for the 
case~a) of black triangles.

In the case~b) we interchange the position of $n_1$ and~$n_2$. 
In the case~c) we require that $\varphi$ depends only on 
$n_1+n_2$. For example, in case~c$'$) we put
$$
2K_2(n)=l_{11}n_1^2+2l_{12}n_1n_2+(l_{22}-l_{21}+l_{12})n_2^2
$$
and make the substitution
$$
\chi_n=w^{n_1}\varphi_{n_1+n_2}.
$$
For the function $\varphi_n$ we obtain the following equation:
$$
\varphi_n+(cwe^{-l_{11}/2}+de^{(l_{21}-l_{12}-l_{22})/2}
e^{(l_{21}-l_{12})(n_1+n_2)})\varphi_{n+1}=0.
$$

In the cases~a)--c) for white triangles we proceed 
completely analogously.
\enddemo

Other cases, when the solutions lie in 
$\script L_2(\mathbb Z^2)$, have not been found so far.

The form $K_2(n)$ can be written as follows.

The case of black triangles:
\roster
\Item"a)" $2K_2=\pmatrix l_{11}&l_{12}\\
l_{12}&l_{22}\endpmatrix$;
\Item"b)" $2K_2=\pmatrix l_{11}&l_{21}\\
l_{21}&l_{22}\endpmatrix$;
\Item"c$'$)" $2K_2=\pmatrix l_{11}&l_{12}\\
l_{12}&l_{22}-l_{21}+l_{12}\endpmatrix$;
\Item"c$''$)" $2K_2=\pmatrix l_{11}-l_{21}+
l_{12}&l_{21}\\l_{21}&l_{22}\endpmatrix$.
\endroster

The case of white triangles:
\roster
\Item"a)" $2K_2=\pmatrix -l_{11}&-l_{12}\\-l_{12}&-l_{22}\endpmatrix$;
\Item"b)" $2K_2=\pmatrix -l_{11}&-l_{21}\\-l_{21}&-l_{22}\endpmatrix$;
\Item"c$'$)" $2K_2=\pmatrix -l_{11}&-l_{12}\\
-l_{12}&-l_{22}+l_{21}-l_{12}\endpmatrix$;
\Item"c$''$)" $2K_2=\pmatrix -l_{11}+l_{21}-l_{12}&
-l_{21}\\-l_{21}&-l_{22}\endpmatrix$.
\endroster

\remark{Remark~$11$}
We can impose analogous conditions on the coefficients 
$l_{ij}$ and their relation with $\alpha,\beta,\delta$, 
demanding that the coefficients of equation (106) for 
$\chi_n$ be dependent only on a combination of the form 
$(\kappa n_1+\tau n_2)$. If $(\kappa n_1+\tau n_2)$ and 
$(un_1+vn_2)$ are the basis of the lattice, then the 
condition of dependency on only one variable has the form:
$$
\aligned
(l_{11}-2\alpha)\tau&=(l_{12}-2\beta)\kappa,
\\
(l_{21}-2\beta)\tau&=(l_{22}-2\delta)\kappa.
\endaligned
\tag 111
$$
Now we can seek a solution in the form
$$
\chi_n=w^{un_1+vn_2}\varphi_{\kappa n_1+\tau n_2}.
\tag 112
$$
Further, for the functions $\varphi_s$, $s\in\mathbb Z$, 
we obtain a difference equation of the form
$$
0=\varphi_s+A(s)\varphi_{s+\kappa}+B(s)\varphi_{s+\tau}.
\tag 113
$$
Only in the three cases mentioned above can this equation 
be easily solved:
$$
\text{a)}\ \ \kappa=1,\ \tau=0;\qquad
\text{b)}\ \ \kappa=0,\ \tau=1;\qquad
\text{c)}\ \ \kappa=\tau=1.
$$
In the other cases we obtain difference equations of order~2 
and higher. We do not know how we could find solutions of them 
that lead to functions $\psi\in\script L_2(\mathbb Z^2)$.
\endremark

>From the theorem we have the following corollary.

\proclaim{Corollary~1}
Under the conditions stated in the theorem the positive 
operators $L=QQ^+$ or $\widetilde L=Q^+Q$ have spectrum 
in $\script L_2(\mathbb Z^2)$ such that the point $\lambda=0$ 
is an infinitely degenerate point of the discrete spectrum.
The eigenfunctions of this point satisfy the equation of 
black or white triangles
$$
1)\ Q\psi=0\quad\text{or}\quad2)\ Q^+\psi=0.
$$
The eigenfunctions of the ground state $\lambda=0$ can 
be determined by explicit formulae, which follow from 
the construction given above.
\endproclaim

This assertion is a difference analogue of results 
in~\cite{7}, \cite{8},~\cite{36} on the ground states 
of the Pauli operator in a continuous purely magnetic case.

\remark{Problems}
1. Prove that the constructed spaces of solutions of 
the equations $Q\psi=0$ and $Q^+\psi=0$ are complete 
in $\script L_2(\mathbb Z^2)$ among all solutions, and 
find appropriate orthonormal bases. Explain what kind 
of cases of solubility in $\script L_2(\mathbb Z^2)$ of 
the equations of black or white triangles exist, apart 
from the solutions found above. Even in the class of 
operators whose coefficients are exponents of linear 
forms $l_j(n)$, this problem has not been solved.

2. Prove that if the conditions of the theorem are satisfied, 
then the spectrum of the operators $L=QQ^+$ and 
$\widetilde L=Q^+Q$ has a non-trivial gap $\Delta$ 
between the ground state $\lambda_0=0$ and the next 
level $\lambda_1\ge\Delta>0$.

3. Find the class of operators $Q$ such that the spectrum 
of the operators $L= QQ^+$ and $\widetilde L=Q^+Q$ 
begins at the point $\lambda=0$ and is continuous, that is, 
the point $\lambda=0$ is the bottom of the continuous 
spectrum ($\Delta= 0$).
\endremark

The last problem is of special interest. Below we consider 
algebraic conditions of the type of commutation of the 
operators $Q$,~$Q^+$, which supposedly lead to the situation 
discussed in Problem~3.

To begin with we consider the difference (purely real) 
$q$-analogues of the continuous Schr\"odinger--Landau 
operator in a homogeneous (constant) magnetic field.

Suppose that the operators $Q$, $Q^+$ have the form~(105):
$$
Q_{c,d}=1+ce^{l_1(n)}T_1+de^{l_2(n)}T_2,
$$
and depend on the parameters
$$
c=e^{k_1},\qquad d=e^{k_2},\qquad l_j=l_{j1}n_1+l_{j2}n_2.
$$

\proclaim{Lemma~16}
Operators $Q$, $Q^+$ of the form $(105)$ satisfy the relation
$$
Q_{c,d}Q_{c,d}^+-1=q(Q_{c',d'}^+Q_{c',d'}-1)
\tag 114
$$
if the matrix $l_{ij}$ and the parameters $c$, $d$, $c'$, $d'$ 
are related by the following equalities\rom:
$$
c=u^{-2}c',\qquad d=u^{-2}d',\qquad q=u^{-2},\qquad u=e^{l_{11}},
$$
$$
2l_{11}=2l_{22}=l_{12}+l_{21}.
\tag 115
$$
\endproclaim

\remark{Remark~$12$}
In the notation
$$
e^{l_{11}}=u,\quad e^{l_{12}}=v
$$
the operator $Q$ is written in the form
$$
Q_{c,d}=1+cu^{n_1}v^{n_2}T_1+d(u^2/v)^{n_1}u^{n_2}T_2.
\tag 116
$$
Dependence on the parameters $(u,v)$ in (114) is omitted. 
Such notation was used in~\cite{43}, where these operators 
were introduced for the first time and their eigenfunctions 
were found.
\endremark

The proof of the lemma is by direct verification.

Comparing the results of Lemma~16 and Theorem~4, we obtain 
the following result~(\cite{43}).

\proclaim{Theorem~5}
We consider the operators $L=QQ^+$ and $\widetilde L=Q^+Q$ 
under the conditions~$(115)$. The spectrum of these operators 
in $\script L_2(\mathbb Z)$ for $0\le\lambda<1$ is purely 
discrete, infinitely degenerate, and can only lie at the 
following points\rom:
$$
\text{\rom{a)}}\ \ \lambda_j=1-u^{2j},\ j\ge0,\quad u<1;
\tag 117
$$
$$
\text{\rom{b)}}\ \ \lambda_j=1-u^{-2j},\ j\ge0,\quad u>1.
\tag 118
$$

In the following cases the spectrum of the operator $L$ 
occupies all points~$(117)$, and the spectrum of the 
operator $\widetilde L$ occupies all points $(117)$ 
except~$\lambda_0$\rom:
$$
\text{\rom{a$'$)}}\ \ u^{-3}>v^{-1}>u^{-1}>1;\qquad
\text{\rom{a$''$)}}\ \ u^{-1}>\max(v,v^{-1})\ge1.
\tag 119
$$

Analogously, the spectrum of the operator $\widetilde L$ 
occupies all points~$(118)$, and $L$ occupies all points 
except~$\lambda_0$, if one of the following conditions 
is satisfied\rom:
$$
\text{\rom{b$'$)}}\ \ u^3>v>u>1;\qquad
\text{\rom{b$''$)}}\ \ u>\max(v,v^{-1})\ge1.
\tag 120
$$

All eigenfunctions discussed in the theorem are obtained 
from solutions of the equations $Q\psi=0$ or $Q^+\psi=0$ 
for proper values of the constants $c$,~$d$,~$u$,~$v$ by 
the use of `creation operators'\rom:

$1)$ $Q_{c,d}\widetilde\psi_0=0$,
$\widetilde L_{c,d}\widetilde\psi_0=Q_{c_0,d_0}^+
Q_{c_0,d_0}\widetilde\psi_0=0$,
$\widetilde\psi=Q_{c_0,d_0}^+\circ\dots\circ 
Q_{c_{k-1},d_{k-1}}^+\widetilde\psi_0$,
$\widetilde L_{c_0,d_0}\psi=(1-u^{\pm2k})\psi;$

$2)$ $Q^+\psi_0=0$, $L_{c,d}\psi_0=Q_{c,d}Q_{c,d}^+\psi_0=0$,
$\psi=Q_{c_0,d_0}\circ\dots\circ Q_{c_{k-1},d_{k-1}}\psi_0$,
$L_{c_0,d_0}\psi=(1-u^{\pm2k})\psi$,
$$
c_j=u^2c_{j-1},\qquad d_j=u^2d_{j-1}.
$$
\endproclaim

\remark{Remark~$13$}
Of the two cases mentioned in the proof of Theorem~4, 
in the given situation only case~1 is realized, in 
which the eigenfunctions $\psi^{(q)}$ that we found 
vanish on the whole subplane.
\endremark

The situation described in Theorem~5 is completely 
analogous to~\cite{22}, that is, to Theorem~1 in~\S3. 
The difference is that for $n=2$ the spaces of solutions 
are infinite-dimensional (the completeness of the solutions 
found on the corresponding levels has not been established 
so far).

We note that conditions (119),~(120) are obtained directly 
from the condition of Theorem~4, but the case~c) here is not 
realized under conditions~(115).

\remark{Problem}
Prove that the spectrum of the operators $L=Q_{c,d}Q_{c,d}^+$ 
and $\widetilde L=Q_{c,d}^+Q_{c,d}$ in the space 
$\script L_2(\mathbb Z^2)$ for $\lambda\ge1$ is continuous 
and runs over the whole semi-axis $\lambda\ge1$. As we assume, 
this is a simple Lebesgue spectrum.
\endremark

We point out that our operators $Q$ depend on the constants 
$c$,~$d$,~$u$,~$v$ (if the dependence on $u$,~$v$ is not 
written explicitly, then in the given formula we assume 
that $u$,~$v$ have not changed). The following formulae are true:
$$
\begin{aligned}
T_1Q_{c,d}T_1^{-1}&=Q_{c',d'},&\qquad
c'&=cd^{l_{11}},&\quad d'&=de^{l_{12}},
\\
T_2Q_{c,d}^+T_2^{-1}&=Q_{c'',d''},&\qquad
c''&=cd^{l_{12}},&\quad d''&=de^{l_{22}},
\end{aligned}
\tag 121
$$
where
$$
\begin{aligned}
Q_{c,d}&=1+ce^{l_1(n)}T_1+de^{l_2(n)}T_2,
\\
Q_{c,d}^+&=q+ce^{l_1(n-T_1)}T_1^{-1}+de^{l_2(n-T_2)}T_2^{-1}.
\end{aligned}
$$
The same is also true for the operators $L=QQ^+$, $\widetilde L=Q^+Q$:
$$
\begin{aligned}
T_1L_{c,d}T_1^{-1}&=L_{c',d'},
\\
T_2L_{c,d}T_2^{-1}&=L_{c'',d''}.
\end{aligned}
$$

\definition{Definition~11}
A {\it characteristic vector-section} is a function 
$\psi_{c,d}(n)$ that is analytically dependent on all 
the parameters $c$,~$d$,~$u$,~$v$ and such that
$$
L\psi=\lambda\psi\quad\text{or}\quad\widetilde L\psi=\lambda\psi,
$$
where $\lambda$ does not depend on the parameters. 
A {\it Bloch vector-section} $\psi_{c,d}(n,p)$ is a 
function $\psi$ such that
$$
L\psi=\lambda\psi,\qquad T_1\psi_{c,d}=e^{ip_1}\psi_{c',d'},\qquad
T_2\psi_{c,d}=e^{ip_2}\psi_{c'',d''}.
$$
\enddefinition

By (121) the functions $T_j\psi_{c,d}$ satisfy these 
equations, but with shifted parameters $(c',d')$ or $(c'',d'')$. 
This definition makes sense also in the one-dimensional 
case~(54), where we had
$$
\gather
Q_c=1+ce^{l(n)}T,\qquad l(n)=ln,
\\
T=T_1,\quad n=n_1,\quad T_1Q_cT_1^{-1}=Q_{ca},\quad a=e^l.
\endgather
$$

Let us construct the Bloch vector-section for $\lambda=0$ 
when $a>1$. We choose $c=c_0>0$. Choosing an arbitrary 
$\psi_{0,c_0}$, we put
$$
\psi_{c_0a^m}(n,p)=T_1^m\bigl(\psi_{0,c_0}(n)\bigr)e^{-imp}.
\tag 122
$$
Limitations on the construction appear as $a\to1$. 
There are no solutions at all in $\script L_2(\mathbb Z)$ 
for $a<1$ if we are talking about the equation $Q_c\psi=0$ 
corresponding to $\lambda=0$. Therefore, $a\ge1$, $c\ne0$.

As $a\to1+0$ the operator tends to an operator with 
constant coefficients, where there are singularities 
of the Bloch vector-section.

Apparently, as $|c|\to1,a\to1$ we can talk about the 
transition of the Bloch vector-section corresponding 
to $\lambda\le1$ into an ordinary Bloch solution for 
the equation with constant coefficients (that is, a 
solution of pure exponent type), since for $c\ne0$ the 
spectrum of the operator for $a=1$ has the form 
$(1-|c|)^2\le\lambda\le(1+|c|)^2$.

In addition, for $a=1$ the spectrum is bounded in 
$\script L_2$, $|\lambda|\le(1+|c|)^2$, and for $a>1$ 
the operator is of course unbounded. Hence, we necessarily 
have a singularity here. For $\lambda\ge1$ the situation 
is not clear.

\remark{Problem}
For the operators $L=QQ^+$ or $\widetilde L=Q^+Q$, whose 
coefficients are exponents of linear functions, construct 
a complete basis of eigenfunctions in the form of 
`Bloch sections' that depend analytically on all the 
parameters $c$,~$d$,~$l_{ij}$, which has minimally possible 
singularities in the submanifolds, where we have operators 
with constant coefficients in one of the variables.
\endremark

We return now to the case when the point $\lambda=0$ is 
supposedly the lower bound of the continuous spectrum 
for operators of the form
$$
L=QQ^+,
\tag 123
$$
where the $Q$ have the form~(105).

\remark{Hypothesis}
If the operators $Q$, $Q^+$ commute, $QQ^+=Q^+Q$, and 
they have a general solution $Q\psi=0$, $Q^+\psi=0$ 
that is bounded on the lattice~$\mathbb Z^2$ (or is 
growing sufficiently slowly?), then the point $\lambda=0$ 
is the lower bound of the continuous spectrum in $\script 
L_2(\mathbb Z^2)$ for the operator $L=\widetilde L$. We 
state below a natural generalization of this hypothesis.
\endremark

\example{Example~10}
Let the operators $Q_{c,d}$, $Q_{c,d}^+$ be such that 
$l_{11}=l_{22}=0$ or $u=1$. Then we have
$$
Q_{c,d}Q_{c,d}^+=Q_{c,d}^+Q_{c,d},
\tag 124
$$
as follows from (114). The operators have the form
$$
\aligned
Q_{c,d}&=1+cv^{n_2}T_1+dv^{-n_1}T_2,
\\
Q_{c,d}^+&=1+cv^{n_2}T_1^{-1}+dv^{-n_1}T_2^{-1}.
\endaligned
\tag 125
$$
\endexample

We pose the following question. Let two operators be 
given in the form
$$
Q_1=1+a_nT_1+b_nT_2,\qquad Q_2=1+c_nT_1^{-1}+d_nT_2^{-1}
\tag 126
$$
with non-vanishing coefficients. We consider the system 
of equations of black and white triangles simultaneously:
$$
\cases
Q_1\psi=0,
\\
Q_2\psi=0.
\endcases
\tag 127
$$
In this case is the system (127) completely locally consistent?

Richer in content is the following formulation. Let 
the `initial condition' for (127) be given in the 
form of two arbitrary values of $\psi$ at the ends of 
any edge of the lattice (for example, $\psi_{n-T_1}$ 
and~$\psi_n$). Using equations~(127), we can solve the 
system and find the value at any other point $n'\in\mathbb Z^2$, 
moving along the paths of the triangles, black and white: 
knowing the values at the ends of any edge, we find the value 
at the third vertex of any black or white triangle that has 
this edge as its border. Considering a bundle of six triangles 
with a common vertex~$n$, we can pass from one cycle and return 
again to the edge $[n-T_1,n]$, and the linear `transformation 
of curvature' will be given by an upper triangular 
$(2\times2)$-matrix:
$$
\gather
\psi_{n-T_1}^{(new)}=A_n\psi_{n-T_1}^{(old)}+B_n\psi_n^{(old)},
\\
\psi_n^{(new)}=\psi_n^{(old)}.
\endgather
$$
The matrix coefficients are the `curvatures' $A_n$, 
$B_n$. They are expressed in terms of the coefficients of 
the system~(127):
$$
\aligned
A_n&=\frac{b_{n-T_1}d_{n+T_1}}{b_nc_{n+T_2}d_na_{n-T_2}},
\\
B_n&=\frac{b_{n-T_1}}{c_{n+T_2}}\biggl(\frac{1}{b_n}\biggl(
\frac{d_{n+T_1}}{a_{n-T_2}}\biggl(\frac{1}{d_n}-b_{n-T_2}\biggr)
-c_{n+T_1}+1\biggr)-d_{n+T_2}\biggr)-a_{n-T_1}.
\endaligned
\tag 128
$$

\proclaim{Lemma~17}
The condition $A_n=1$, $B_n=0$ for all $n\in\mathbb Z^2$ 
is necessary and sufficient for the possibility of a unique 
solution of the system $(127)$ on the plane $\mathbb Z^2$ 
under arbitrary initial conditions imposed at the ends of 
any fixed edge. For this it is also necessary and sufficient 
to satisfy the following algebraic relation for the 
operators $Q_1$,~$Q_2$\rom:
$$
\bigl((Q_1-1)(Q_2-1)-1\bigr)=f_n\bigl((Q_2-1)(Q_1-1)-1\bigr),\qquad
f_n\ne0.
\tag 129
$$
\endproclaim

We omit the proof of the lemma.

\proclaim{Corollary~2}
If the system $(127)$ is completely consistent, then its 
solution is determined by an arbitrary solution of a 
one-dimensional difference equation of the second order 
along any path of edges without self-crossing, in particular, 
along the path $n_1=\const$ \rom(we obtain an equation for 
the variable~$n_2)$ or $n_2=\const$ \rom(we obtain an equation 
for the variable~$n_1)$. If the curvature is trivial, then all 
these equations are equivalent.
\endproclaim

\example{Example~11}
For the commuting operators $Q_1=Q_2^+=Q_{c,d}$ from (125) 
the condition for consistency is satisfied. For the 
variable~$n_1$ ($n_2=\const$) we get the equation
$$
cv^{n_2}(\psi_{n-T_1}+\psi_{n+T_1})+(1+c^2v^{2n_2}-
dv^{-2n_1})\psi_n=0.
\tag 130
$$
\endexample

To end this section we give some information on factorizations 
and Laplace transformations for general non-self-adjoint 
operators $L$ of the second order on an equilateral triangular 
lattice, and also on complex Hermite operators. The following 
assertions hold.

\proclaim{Theorem~6}
To represent the operator
$$
L=a_n+b_nT_1+c_nT_2+d_{n-T_1}T_1^{-1}T_2+e_{n-T_1}T_1^{-1}+
f_{n-T_2}T_2^{-1}+g_{n-T_2}T_1T_2^{-1}
\tag 131
$$
in the form
$$
\text{\rom{a)}}\ \ L=Q_1Q_2+w_n\quad\text{or}\quad
\text{\rom{b)}}\ \ L=Q_2Q_1+w_n,
$$
where $Q_1$ and $Q_2$ have the form
$$
Q_1=x_n+y_nT_1+z_nT_2,\qquad
Q_2=p_n+q_{n-T_1}T_1^{-1}+r_{n-T_2}T_2^{-1},
$$
it is necessary and sufficient to satisfy the conditions
$$
\begin{aligned}
\text{\rom{a)}}&\ \ b_{n+T_2}d_nf_{n+T_1}=g_ne_{n+T_2}c_{n+T_1},
\\
\text{\rom{b)}}&\ \ f_nd_nb_n=c_ne_ng_n.
\end{aligned}
\tag 132
$$
If both conditions are satisfied, then the Laplace transformations
$$
\text{\rom{a)}}\ \ L=Q_1Q_2+w\longmapsto\widetilde L=Q_2w^{-1}Q_1+1,
\tag 133
$$
$$
\text{\rom{b)}}\ \ L=Q_2Q_1+v\longmapsto\widetilde L=Q_1w^{-1}Q_2+1
\tag 134
$$
can be iterated, moreover infinitely many times \rom(if~$w,v\ne0)$.
\endproclaim

\remark{Remark~$14$}
The factorization conditions (132) are invariant with respect 
to the rotation of the lattice by the angle~$2\pi/3$. By 
analogy with Definition~10 we can introduce six types of 
Laplace transformations. Then the 1st, 3rd and 5th 
transformations are unitarily adjoint and inverse to the 
4th, 6th and 2nd transformations respectively. The same applies 
to the following assertion.
\endremark

\proclaim{Theorem~7}
To factorize an operator $L$ of the form
$$
L=a_n+b_nT_1+\overline c_nT_2+d_{n-T_1}T_2T_1^{-1}
+\overline b_{n-T_1}T_1^{-1}+c_{n-T_2}T_2^{-1}+
\overline d_{n-T_2}T_1T_2^{-1}
\tag 135
$$
in the form
$$
\text{\rom{a)}}\ \ L=QQ^++w_n\quad\text{or}\quad
\text{\rom{b)}}\ \ L=Q^+Q+w_n,
$$
it is necessary and sufficient to satisfy the following 
conditions\rom:
$$
\aligned
\text{\rom{a)}}&\ \ b_{n-T_1}c_{n-T_2}d_{n-T_1-T_2}\in\mathbb R,
\\
\text{\rom{b)}}&\ \ d_nc_nb_n\in\mathbb R.
\endaligned
\tag 136
$$
In particular, if both conditions are satisfied, then the 
Laplace transformations can be iterated infinitely many times.
\endproclaim

The proof of these theorems is straightforward.

For complex Hermite operators $L$ it is natural to define 
a class of `phase' gauge transformations 
$$
L\mapsto e^{if_n}Le^{-if_n},\qquad
\psi_n\mapsto e^{if_n}\psi_n
$$
such that $f_n\in\mathbb R$. These transformations keep 
the operator formally Hermitian with respect to the 
previous standard scalar product in~$\script L_2(\mathbb Z^2)$
$$
\langle\varphi,\psi\rangle=\sum_{n\in\mathbb Z^2}
\varphi_n\overline\psi_n.
\tag 137
$$
The operator $L$ naturally has `real' and `phase' projections
$$
\begin{aligned}
L_{\mathbb R}&=a_n+\beta_nT_1+\gamma_nT_2+\delta_{n-T_1}T_2T_1^{-1}
+\beta_{n-T_1}T_1^{-1}+\gamma_{n-T_2}T_2^{-1}+\delta_{n-T_2}T_1T_2^{-1},
\\
L_\Phi&=a_n\,{+}\,B_nT_1\,{+}\,\overline 
C_nT_2\,{+}\,D_{n-T_1}T_2T_1^{-1}
\,{+}\,\overline B_{n-T_1}T_1^{-1}\,{+}\,
C_{n-T_2}T_2^{-1}\,{+}\,\overline D_{n-T_2}T_1T_2^{-1},
\end{aligned}
$$
$b_n=\beta_nB_n$, $c_n=\gamma_nC_n$, $d_n=\delta_nD_n$, 
where $\beta_n$, $\gamma_n$, $\delta_n$, $a_n$,~$w_n$ are 
real numbers and $B_n$,~$C_n$,~$D_n$ are complex with modulus 
equal to~1.

\definition{Definition~12}
A {\it physical magnetic field} is the phase part of 
the product of the coefficients of an operator (of connection) 
along the border of any black or white triangle
$$
e^{-i\Phi_n^{(1)}}=B_{n-T_1}C_{n-T_2}D_{n-T_1-T_2},
\tag black
$$
$$
e^{i\Phi_n^{(2)}}=D_nC_nB_n.
\tag white
$$
\enddefinition

The following lemma is obvious.

\proclaim{Lemma~18}
The operator $L$ is reduced by the phase gauge 
transformation \linebreak$e^{if_n}Le^{-if_n}$ to a purely 
real form if and only if the magnetic field \rom(that is, 
all `magnetic fluxes' $\Phi_n^{(j)})$ is trivial\rom:
$$
\Phi_n^{(j)}=0,\qquad n\in\mathbb Z^2,\quad j=1,2.
$$
\endproclaim

\proclaim{Corollary~3}
A complex Hermite operator $L$ admits an unbounded 
number of Laplace transformations of all six forms 
if and only if it is reduced by a phase gauge transformation 
to a purely real operator.
\endproclaim

Thus, the $q$-analogues of the Schr\"odinger--Landau 
operator on the lattice have no connection with the 
magnetic field. Difference operators in a physical 
magnetic field do not factorize, generally speaking. 
A~proper analogue of an operator in a homogeneous 
magnetic field can (having chosen a specific gauge 
transformation, that is, a `vector-potential' composed 
of coefficients $b_n,c_n,d_n$ such that $|b_n|=|c_n|=|d_n|=1$) 
be written in the form
$$
\begin{aligned}
L=6-e^{i\Phi n_2}T_1-e^{-i\Phi n_1}T_2-
e^{i\Phi(n_1+n_2)}T_2T_1^{-1}
\\
-e^{-i\Phi n_2}T_1^{-1}-e^{i\Phi n_1}T_2^{-1}-
e^{-i\Phi(n_1+n_2)}T_1T_2^{-1}.
\end{aligned}
\tag 138
$$

\head
\boldsymbol\S7. Factorizations and Laplace transformations 
on many-dimensional lattices of regular tetahedra 
in~$\mathbb R^{\bf N}$
\endhead

We consider a lattice in $\mathbb R^N$ such that the ends 
of the basis vectors $T_1,\dots,T_N$ of the lattice together 
with the point~0 form a regular $N$-dimensional simplex 
(a~tetrahedron for $N=3$). We call this simplex, and also 
all others obtained from it through translations by a 
vector of the lattice, {\it black} tetrahedra. {\it White} 
tetrahedra ($N$-dimensional simplexes) are the tetrahedron 
$\langle0,-T_1,\dots,-T_N\rangle$ and all others obtained 
from it through integer-valued translations.

Two black (white) tetrahedra can have no more than one 
vertex in common. A~black and a white tetrahedron can 
have the longest edge in common. This is true for any 
dimension~$\ge2$.

We consider a real self-adjoint operator of the form
$$
L=a_n+\sum_{k=1}^N(b_{k,n}T_k+b_{k,n-T_k}T_k^{-1})+
\sum_{k\ne j}(c_{kj,n-T_j}T_kT_j^{-1}+c_{kj,n-T_k}T_jT_k^{-1}).
$$
We assume that the coefficients $b_{k,n},c_{kj,n}$ are a 
real connection, and they are all non-zero.

\proclaim{Theorem~8}
The operator $L$ admits a representation in the form
$$
L=QQ^++w_n,
$$
where
$$
Q=x_n+\sum_{k=1}^Ny_{k,n}T_k,
$$
if and only if the following condition for the coefficients 
$b_{k,n}$,~$c_{kj,n}$ is satisfied\rom:
$$
\frac{b_{k,n-T_k}b_{j,n-T_j}}{c_{kj,n-T_k-T_j}}=x_n
\text{ does not depend on $k$, $j$}.
\tag 139
$$
Analogously, the condition for factorization $L=Q^+Q+w_n$ 
is the independence of the expression $b_{k,n}b_{j,n}/c_{kj,n}$ 
on~$k$,~$j$.
\endproclaim

\remark{Remark~$15$}
For $N=2$ we do not have any conditions. For $N=3$ we have 
two conditions for any white tetrahedron: the product of the 
coefficients of connection that sit on pairs of opposite 
(skew) edges of the tetrahedron is the same. In all, the 
tetrahedron has six edges, which make up three skew pairs, 
hence we have two conditions. If we want to have all forms 
of factorizations, then this condition should also be satisfied 
for black tetrahedra.
\endremark

The proof of the theorem, as before, is by direct 
verification.

We now consider factored operators, for which the 
coefficients of the corresponding operators $Q$,~$Q^+$ 
have the form of exponents of linear functions, as for~$N=2$:
$$
Q=1+ce^{l_1(n)}T_1+de^{l_2(n)}T_2+fe^{l_3(n)}T_3,
\tag 140
$$
$$
l_j(n)=\sum_{i=1}^Nl_{ji}n_i.
$$
For such operators the following results are true.

\proclaim{Theorem~9}
The equations
$$
\text{\rom{1)}}\ \ Q\psi=0\text{ \rom(the equation of 
black triangles\rom)}
\tag 141
$$
or
$$
\text{\rom{2)}}\ \ Q^+\psi=0\text{ \rom(the equation of 
white triangles\rom)}
\tag 142
$$
certainly have infinite-dimensional spaces of solutions 
$\psi\in\script L_2(\mathbb Z^N)$ if the following conditions 
are satisfied\rom:

$1)$
$$
\text{\rom{1a)}}\ \ \pmatrix
l_{11}&l_{12}&l_{13}\\
l_{12}&l_{22}&l_{23}\\
l_{13}&l_{23}&l_{33}
\endpmatrix>0,\enskip l_{32}=l_{23},
\tag 143
$$
$$
\text{\rom{1b)}}\ \ \pmatrix
l_{11}&l_{12}&l_{13}\\
l_{12}&l_{22}-l_{21}+l_{12}&l_{23}\\
l_{13}&l_{23}&l_{33}
\endpmatrix>0,\enskip l_{23}+l_{31}=l_{32}+l_{13},
\tag 144
$$
$$
\text{\rom{1c)}}\ \ \pmatrix
l_{11}&l_{12}&l_{13}\\
l_{12}&l_{22}-l_{21}+l_{12}&l_{23}-l_{21}+l_{12}\\
l_{13}&l_{23}-l_{21}+l_{12}&l_{33}-l_{31}+l_{13}
\endpmatrix>0,\enskip l_{23}-l_{21}+l_{12}=l_{32}-l_{31}+l_{13};
\tag 145
$$

$2)$ the same with $l_{ij}\to-l_{ij};$

\noindent
and also the conditions obtained from those stated above 
by an arbitrary permutation of the indices~$1,2,3$.
\endproclaim

\proclaim{Corollary~4}
If the conditions of the theorem are satisfied, the 
operators $L=QQ^+$ and $\widetilde L=Q^+Q$ have $\lambda=0$ 
as the point of the discrete spectrum that is infinitely 
degenerate.
\endproclaim

\remark{Problem}
Prove that the eigenfunctions constructed according to the 
scheme of the proof of Theorem~4 give a complete basis 
for~$\lambda=0$. Prove that the remaining spectrum is 
separated from $\lambda=0$ by a finite gap $\Delta>0$ 
in the space $\script L_2(\mathbb Z^3)$.
\endremark

\proclaim{Theorem~10}
If the relations
$$
l_{ij}+l_{ji}=h
\tag 146
$$
are satisfied, then the operators $Q$, $Q^+$ satisfy the relations 
$$
Q_{c,d,f}Q_{c,d,f}^+-1=q(Q_{c',d',f'}^+Q_{c',d',f'}-1),
\tag 147
$$
where
$$
q=e^h,\qquad c'=e^{-h}c,\qquad d'=e^{-h}d,\qquad f'=e^{-h}f.
$$
\endproclaim

This theorem generalizes Lemma~16, formulated for~$N=2$.

It is essential to note that the conditions (146) contradict 
the condition of \linebreak Theorem~9, and we cannot 
explicitly find solutions of the equations
$$
Q_{c,d,f}\psi =0,
\tag 148
$$
$$
Q_{c,d,f}^+\psi=0,
\tag 149
$$
that belong to $\script L_2(\mathbb Z^2)$ and are necessary 
to construct the spectrum of the operators $L$,~$\widetilde L$.

\remark{Remark~$16$}
As before, we can seek solutions of (141) in the form
$$
\psi_n=e^{-K_2(n)}\chi_n,
$$
where $K_2(n)$ is chosen in such a way that the coefficients 
of the equation for $\chi_n$ do not contain the variable~$n_1$. 
However, we have not succeeded in reducing the resulting 
equation to one variable. After the substitution
$$
\chi_n=w^{n_1}\varphi_{n_2,n_3}
$$
we arrive at the two-dimensional equation of triangles
$$
A_n\varphi_n+B_n\varphi_{n+T_2}+C_n\varphi_{n+T_3}=0.
\tag 150
$$
Unfortunately, the coefficients of this equation are very 
complicated, and so far we do not know how to find solutions 
of it explicitly.
\endremark

Obviously, one equation (141) can always be solved in the 
half-space of a definite direction if the initial conditions 
are imposed arbitrarily on vertices in any plane that has 
one of the four forms:
$$
n_1=\const,\ \text{or}\ n_2=\const,\ \text{or}\ n_3=\const,\
\text{or}\ n_1+n_2+n_3=\const.
\tag 151
$$
Here the equation of a tetrahedron (as of a triangle for $N=2$) 
is treated as the evolution equation. It is not reversible, 
as it is for $N=2$, since a solution in the inverse direction 
of `time' is no longer local; its solubility depends on the 
initial condition belonging to a special functional class.

Now we consider the condition for consistency of the pair of 
equations (black and white)
$$
\aligned
Q_1\psi&=0,\qquad Q_1=1+x_nT_1+y_nT_2+z_nT_3,
\\
Q_2\psi&=0,\qquad Q_2=1+p_nT_1^{-1}+q_nT_2^{-1}+r_nT_3^{-1}.
\endaligned
\tag 152
$$

\proclaim{Theorem~11}
The system $(152)$ is completely consistent, that is, its 
solution is determined by an arbitrary solution of some 
difference equation of the second order in one of the 
planes~$(151)$, if the relation
$$
(1-Q_1)(1-Q_2)-1=f_n((1-Q_2)(1-Q_1)-1)
\tag 153
$$
is satisfied, where $f_n$ is a non-zero function \rom(in 
particular, for $f_n\equiv1$ this equality is transformed 
into the commutation condition $Q_1Q_2=Q_2Q_1)$. The 
corresponding equation of the second order, for example 
for the plane $n_1+n_2+n_3=\const$, has the form
$$
(1-Q_1)(1-Q_2)\psi=\psi.
$$
\endproclaim

\demo{Proof}
Since equations (152) have the form
$$
(1-Q_1)\psi=\psi,\qquad (1-Q_2)\psi=\psi,
$$
we see that the conditions
$$
(1-Q_1)(1-Q_2)\psi=\psi
$$
and
$$
(1-Q_2)(1-Q_1)\psi=\psi,
$$
which are equations of the second order in the plane 
$n_1+n_2+n_3=\const$, should be equivalent. Hence it 
follows that there is a non-zero function $f_n$ such that
$$
(1-Q_1)(1-Q_2)-1=f_n\bigl((1-Q_2)(1-Q_1)-1\bigr).
$$
Thus, we have proved the theorem for the plane $n_1+n_2+n_3=\const$.

Writing our conditions locally, in the star of any vertex, 
we see that it is invariant with respect to rotations that 
transform the lattice into itself. All four directions of 
the planes mentioned above are equivalent. We have proved 
Theorem~11.
\enddemo

\example{Example~12}
Let us consider the operators~(147), where $h=0$. In this 
case the operators $Q$,~$Q^+$ commute. The corresponding 
equation in the hyperplane $n_3=\const$ has the form
$$
\begin{aligned}
0=(A+C^2e^{2\alpha n_2}+D^2e^{-2\alpha n_1})\psi_n+
Ce^{\alpha n_2}(\psi_{n+T_1}+\psi_{n-T_1})
\\
\hphantom{....}+De^{-\alpha n_1}(\psi_{n+T_2}+\psi_{n-T_2})+
CDe^{\alpha(n_2-n_1)}(e^\alpha\psi_{n+T_2-T_1}
+e^{-\alpha}\psi_{n+T_1-T_2}).
\end{aligned}
\tag 154
$$
\endexample

Now we consider a `vector factorization' for $N=3$, where 
the operators $Q^+$ represent vectors $(Q^{+\alpha})$, 
$\alpha=1,2$. Hence, the factorization of the operator 
is given in the form
$$
\aligned
\text{\rom{a)}}&\ \ \text{\rom{WW}}\:\
L=\sum_{\alpha=1}^2Q^\alpha Q^{+\alpha}+w_n,
\\
\text{\rom{b)}}&\ \ \text{\rom{BB}}\:\
L=\sum_{\alpha=1}^2Q^{+\alpha}Q^\alpha+w_n,
\\
\vspace{6pt}
\text{\rom{c)}}&\ \ \text{\rom{BW}}\:\
L=Q^1Q^{+1}+Q^{+2}Q^2+w_n,
\endaligned
\tag 155
$$
where
$$
Q^\alpha=x_n^\alpha+y_n^\alpha T_1+z_n^\alpha T_2+t_n^\alpha T_3.
$$

\proclaim{Lemma~19}
Vector factorization of a self-adjoint operator $L$ is 
always possible \rom(although it is not unique\rom). 
Moreover, we can regard the potential as a constant $w_n=w_0$.
\endproclaim

The vector factorizations of all three forms generate 
Laplace transformations which, however, cannot be iterated.

The search for eigenfunctions of the ground state can 
sometimes be carried out in one of them:
$$
\aligned
\text{a)}&\ \ Q^{+1}\psi=0,\ Q^{+2}\psi=0\text{ for (WW)},
\\
\text{b)}&\ \ Q^1\psi=0,\ Q^2\psi=0\text{ for (BB)},
\\
\text{c)}&\ \ Q^{+1}\psi=0,\ Q^2\psi=0\text{ for (BW)}.
\endaligned
\tag 156
$$
If the constant $w_0$ for the factorization is chosen 
correctly, then the lowest level is obtained from 
equations~(156), a),~b), or~c).

It is difficult to find a criterion for the existence of one 
solution for the systems~(156). However, as before, we state 
here a criterion for complete local consistency of these systems.

\proclaim{Theorem~12}
The system~\rom{(156) a) (}or~\rom{b)} has solutions 
that are uniquely determined by arbitrary initial conditions 
imposed on the vertices of any straight line with direction 
vector $T_1$,~$T_2$,~$T_3$ or $T_iT_j^{-1}$,
$i\ne j$, inside the dihedral angle in which lie the 
white \rom(respectively, black\rom) tetrahedra adjoining 
this straight line along an edge, if and only if the 
following condition is satisfied\rom: for any four white 
\rom(respectively, black\rom) tetahedra that are located 
in angles of a tetrahedron twice the size, the eight 
equations corresponding to them are linearly dependent.
\endproclaim

\demo{Proof}
For definiteness we consider the system (156)~b). For the 
values of $\psi$ at the vertices of every black tetrahedron 
we have two equations. This means that if we know the values 
of $\psi$ at two vertices of a black tetrahedron, we can 
find the values at the other two. Hence we can easily see 
that if $\psi$ is given on all the vertices of some straight 
line, then our system dictates the extension of $\psi$ into 
the interior of the stated sector. Thus, we only need to 
clarify when this extension is possible for any initial 
conditions.

We consider a tetrahedron $T_N$, similar to the black one, 
with edge of length~$N$. Let $V(N)$ be the number of black 
tetrahedra that lie in $T_N$,
$V(N)=N(N+ 1)(N+ 2)/6$. 
For the values of $\psi$ at the $V(N+1)$ vertices that lie 
inside and on the border of $T_N$ we have $2V(N)$ equations, 
of which only $V(N+1)-N-1$ should be linearly independent in 
order to satisfy the condition for complete consistency. But 
$2V(N)-V(N+ 1)+N+1=V(N-1)$ is the number of 
double-size tetrahedra $T_2\subset T_N$.
\enddemo

The condition for consistency of the system (156)~c) was 
considered in \linebreak Theorem~11.

\head
\boldsymbol\S8. Factorizations of operators and Laplace 
transformations on two-dimensional surfaces
\endhead

Let us consider a two-dimensional manifold without boundary, 
triangulated in such a way that two-dimensional simplexes can 
be painted in two colours (black and white) so that two 
triangles that are adjacent along an edge have different colours. 
In this case an even number of triangles should meet at each 
vertex.

The metric on the surface is chosen in such a way that all 
triangles are equivalent to an equilateral triangle in the 
Euclidean plane. Hence it follows immediately that at any 
vertex the total curvature has the form $(2\pi-N_{\tr}\pi/3)$, 
where $N_{\tr}$ is the number of triangles meeting at a given 
vertex. If $N_{\tr}=6$, then the curvature is equal to zero. 
If $N_{\tr}=4$, then the curvature is positive. If $N_{\tr}>6$, 
then the curvature is negative.

\definition{Definition~13}
A {\it vertex scalar Schr\"odinger operator} is an operator 
$L$ that acts on a function from a vertex according to the formula
$$
(L\psi)_P=\sum_{P'}b_{P:P'}\psi_{P'}+a_P\psi_P,
\tag 157
$$
where summation goes over the vertices $P'$ that are nearest 
to $P$ but do not coincide with~$P$. The condition that the 
operator $L$ is self-adjoint has the form
$$
b_{P':P}=\overline b_{P:P'},\qquad a_P\in\mathbb R.
\tag 158
$$
\enddefinition

\definition{Definition~14}
1) A {\it triangular \rom(black\rom) Schr\"odinger operator 
of type}~I is an operator $L$ that acts on functions $\psi_T$ 
of the black triangles $T$ in such a way that
$$
(L\psi)_T=\sum_{T'}b_{T:T'}\psi_{T'}+a_T\psi_T,
\tag 159
$$
where $T'$ are all the black triangles that have a common 
vertex with~$T$.

2) A {\it triangular \rom(black\rom) Schr\"odinger operator of 
type}~II is an operator $L$ that acts on $\psi_T$ according to 
formula~(159), but now $T'$ runs over all black triangles that 
have with $T$ a common neighbouring (that is, adjacent along an 
edge) white triangle.

Analogously we define white triangular operators of types~I 
and~II. In both cases the self-adjointness condition with respect 
to the standard scalar product
$$
\langle\varphi,\psi\rangle=\sum_T\varphi_T\overline\psi_T
\tag 160
$$
has the form
$$
b_{T':T}=\overline b_{T:T'},\qquad a_T\in\mathbb R.
$$
\enddefinition

\proclaim{Theorem~13}
Any real self-adjoint vertex operator $L$ admits a unique 
factorization of the form
$$
L=QQ^++w_P,
$$
where
$$
(Q^+\psi)_T=\sum_{P}y_{T:P}\psi_P+x_T\psi_T,
\tag 161
$$
and the sum is taken over all vertices $P$ of the black 
triangle~$T$. For any black triangle $T$ and two of its 
vertices $P_1$,~$P_2$ we have the equality \rom(compare 
with~$(101))$\rom:
$$
y_{T:P_1}y_{T:P_2}=b_{P_1:P_2}.
\tag 162
$$
\endproclaim

We have also an analogous assertion for white triangles.

\proclaim{Theorem~14}
A real triangular black self-adjoint operator $\widetilde L$ 
of type~\rom I admits a factorization of the form
$$
\widetilde L=Q^+Q+v_T,
\tag 163
$$
where the operator $Q^+$ has the form~$(161)$, if and only 
if for every vertex $P$ the matrix $B_P=(b_{T:T'})$, where 
the black triangles $T$,~$T'$ belong to the star of the 
vertex~$P$, has the form
$$
B_P=\operatorname{Diag}+\Lambda_P,\qquad \rk\Lambda_P=1.
\tag 164
$$
For multiplicities $N_{\tr}=4,6$ this condition is always 
satisfied.

The operator $Q$ in $(163)$ is defined up to a transformation 
that does not change any product of the form
$$
y_{T:P}y_{T':P}\qquad (T\ne T').
$$
In particular, for $N_{\tr}\ge6$ the coefficients $y_{T:P}$ 
are defined uniquely, and for $N_{\tr}=4$ up to the 
transformation $(T\ne T')$
$$
y_{T:P}\mapsto\mu_Py_{T:P},\qquad y_{T':P}\mapsto\mu_P^{-1}y_{T':P},
$$
where $0\ne\mu_P$ is any non-zero function from vertices 
with $N_{\tr}(P)=4$.
\endproclaim

\proclaim{Theorem~15}
A triangular \rom(black\rom) Schr\"odinger operator of 
type~\rom{II} always admits a factorization of the form
$$
L=QQ^++u_T,
\tag 165
$$
where
$$
(Q^+\psi)_{T_1}=\sum_Ty_{T_1:T}\psi_T+x_{T_1}\psi_{T_1},
$$
$T_1$ is a white triangle, and the sum is taken over all 
black triangles $T$ adjacent to~$T_1$. In fact, the coefficients 
$y_{T_1:T}$ correspond to the edges of the triangulation.

We have the equalities
$$
y_{T:T_1}y_{T':T_1}=b_{T:T'}
\tag 166
$$
for any three triangles $T$, $T_1$,~$T'$ sequentially adjacent 
to each other \rom(black, white, black\rom) from the star of a 
vertex with $N_{\tr}\ge6;$
$$
y_{T:T_1}y_{T':T_1}+y_{T:T_1'}y_{T':T_1'}=b_{T:T'}
\tag 167
$$
for black triangles $T$, $T'$ and white triangles $T_1$, $T_1'$ 
from the star of a vertex with $N_{\tr}=4$.

The factorization is unique in the neighbourhood of all 
vertices $P$ such that $N_{\tr}(P)\ge 6$. In 
the neighbourhood of vertices $P$ with $N_{\tr}(P)=4$ the 
factorization is not unique\rom: every equation of the form 
$(167)$ can be replaced by two\rom:
$$
y_{T:T_1}y_{T':T_1}=b^{(1)}_{T:T'}\,,\qquad
y_{T:T_1'}y_{T':T_1'}=b_{T:T'}^{(2)}\,,
\tag 168
$$
where $b_{T:T'}=b_{T:T'}^{(1)}+b_{T:T'}^{(2)}$ is an arbitrary 
decomposition of $b_{T:T'}$ into a sum of two non-zero terms. 
The resulting system is uniquely soluble.
\endproclaim

In this theorem the black and white colours can be interchanged.

\remark{Remark~$17$}
If for all vertices of the triangulation we have $N_{\tr}=4,6$, 
then the classes of operators of types~I and~II coincide.
\endremark

\proclaim{Corollary~5}
$1)$ If for all vertices we have $N_{\tr}\le6$ \rom(that is, 
the curvature is non-negative\rom), then factorizations of all 
types are defined. Correspondingly, all the Laplace 
transformations related to them are also defined. For example, 
for a vertex operator the Laplace transformation has the form
$$
L=QQ^++w\longmapsto\widetilde L=Q^+w^{-1}Q+1.
\tag 169
$$

$2)$ Factorizations of triangular Schr\"odinger operators $L$ 
of type~\rom{II} are always defined in terms of triangles of 
the opposite colour. The degree of non-uniqueness of these 
factorizations and the corresponding Laplace transformations 
is determined by the number of vertices $P$ of positive 
curvature, $N_{\tr}(P)=4$.
\endproclaim

\remark{Remark~$18$}
For an equilateral triangular lattice in the plane the spaces 
of functions on vertices, on black and on white triangles, can 
be identified, and then these factorizations coincide (up to 
equivalence~(104)) with the factorizations of operators on a 
regular lattice considered in~\S6.
\endremark

\smallskip\noindent{\bf Conclusion.}
Formulation of the problem on cyclic Laplace chains arises in 
two cases.

\smallskip
\noindent{\sl Case}~1: for $N_{\tr}\le6$ (Corollary~5).

\smallskip
\noindent{\sl Case}~2: for vertices with $N_{\tr}=4$ under the 
conditions of Corollary~5.

\smallskip
In case~2 we can limit ourself to Laplace transformations for 
operators of type~II, which transform functions on black 
triangles into functions on white triangles and vice versa. 
This is similar to one-dimensional Darboux transformations.

\head
\boldsymbol\S9. Simplicial connections. Generalizations
\endhead

We consider a simplicial complex~$K$. Everywhere we denote a 
simplex of dimension $l$ by~$\sigma^l$.

\definition{Definition~15}
A {\it simplicial connection of type} $(q,j,k)$, $0\le j<k$, 
is an equation
$$
\sum_{\sigma^q\subset\sigma^{q+k}}c_{\sigma^{q+k}:\sigma^q}
\psi_{\sigma^q}=0,
\tag 170
$$
determined by a vector-function which ascribes to every pair of 
simplexes \linebreak$\sigma^q,\sigma^{q+k}\subset K$ 
such that $\sigma^q\subset\sigma^{q+k}$ a collection
$$
c_{\sigma^{q+k}:\sigma^q}=(c_{\sigma^{q+k}:\sigma^q}^\alpha)\in\mathbb 
R^m,\qquad
\alpha=1,\dots,m,
\tag 171
$$
where
$$
m=C_{q+k+1}^{q+1}-C_{q+j+1}^{q+1}.
$$
We require that the function $c_{\sigma^{q+k}:\sigma^q}$ 
satisfies the conditions for non-degeneracy and localization 
(see below).
\enddefinition

Such a function defines an operator $Q^+$ which transforms the 
space of numerical functions of simplexes $\sigma^q$ into the 
space of $m$-vector-functions of simplexes~$\sigma^{q+k}$:
$$
(Q^+\psi)_{\sigma^{q+k}}=\sum_{\sigma^q}c_{\sigma^{q+k}:\sigma^q}
\psi_{\sigma^q}\in\mathbb R^m.
\tag 172
$$
Simplicial connection depends only on the zero-space of this 
operator:
$$
Q^{+\alpha}\psi=0,\qquad \alpha=1,\dots,m.
\tag 173
$$
In fact, this relation is written separately in each 
simplex~$\sigma^{q+k}$.

\smallskip
\noindent{\sl Requirement of non-degeneracy}. Equation (197) 
should be such that for any subsimplex $\sigma^{q+k}\subset K$ 
an arbitrarily given collection of values $\psi_{\sigma^q}$ in 
simplexes $\sigma^q$, forming a $q$-dimensional skeleton of any 
simplex $\sigma^{q+j}\subset\sigma^{q+k}$, should uniquely and 
consistently determine values of $\psi$ on all remaining 
$q$-dimensional subsimplexes in $\sigma^{q+k}\subset K$. Thus, 
we can arbitrarily define values of $\psi$ in the $q$-dimensional 
skeleton of simplexes $\sigma^{q+j}\subset\sigma^{q+k}$.

\proclaim{Lemma~20}
Suppose we are given a path $\gamma$ composed of 
$(q+k)$-dimensional simplexes $\sigma_1^{q+k},\sigma_2^{q+k},\dots$, 
where $\sigma_s^{q+k}$ and $\sigma_{s+1}^{q+k}$ intersect 
exactly along a face of dimension $q+ j$ for all~$s$. 
Then the simplicial connection consistently defines a solution 
of $(170)$ along the path~$\gamma$, starting from arbitrary 
initial data defined in a $q$-dimensional skeleton of an arbitrary 
$(q+j)$-dimensional face of any of the simplexes $\sigma^{q+k}_s$ 
that make up this path.
\endproclaim

We obtain the proof of the lemma in a simple way from the 
definitions: a~solution is constructed by transition from 
the simplex $\sigma^{q+k}_s$ to the simplex $\sigma^{q+k}_{s+1}$, 
$s=1,2,\dots$\,.

The analogue of curvature arises naturally for `closed' 
paths $\gamma$ in which \linebreak$\sigma_1^{q+k}=
\sigma_N^{q+k}$. In this case there arises 
a `holonomy transformation' of the simplicial connection 
(170) along the path~$\gamma$: solving (170) along the path 
according to the lemma, starting from some face 
$\sigma^{q+j}_0\subset\sigma_1^{q+k}$, we finally find 
the value of $\psi$ on all $q$-dimensional faces of the 
simplex $\sigma_N^{q+k}$, including $\sigma^q\subset\sigma^{q+j}_0$. 
These values may not coincide with the initial values. There arises 
the linear transformation
$$
R_\gamma\:\mathbb R^M\to\mathbb R^M,
$$
where $M=C_{q+j+1}^{q+1}$ is the number of $q$-dimensional 
faces in the simplex $\sigma_0^{q+j}\subset\sigma_1^{q+k}$ 
on which the values of the function $\psi_{\sigma^q}$ were 
given arbitrarily.

Starting from the requirement of non-degeneracy we can reduce 
equation (170) in any simplex $\sigma^{q+k}$ to such a form that 
inverse operators are defined for any pair of simplexes 
$\sigma_1^{q+j},\sigma_2^{q+j}\subset\sigma^{q+k}$:
$$
L_{12}\:\mathbb R_1^M\to\mathbb R_2^M,
$$
where the space $\mathbb R_\varepsilon^M$ consists of all possible 
values of $\psi$ on simplexes 
$\sigma^q\subset\sigma^{q+j}_\varepsilon$, $\varepsilon=1,2$, 
and the bases in them are $\delta$-functions of the 
simplexes~$\sigma^q$.

\smallskip
\noindent{\sl Requirement of localization}. All operators 
$L_{12}$ are uniquely defined only by the minimal simplex 
containing $\sigma_1^{q+j}$ and~$\sigma_2^{q+j}$:
$$
(\sigma_1^{q+j}\cup\sigma_2^{q+j})\subset\sigma^{q+j+s}\subseteq
\sigma^{q+k}.
$$
This means that if two simplexes $\sigma_1^{q+k}$ and 
$\sigma_2^{q+k}$ intersect along the simplex $\sigma^{q+j+s}$, 
where $s>0$, then for them all the operators $L_{12}$ inside 
the simplex $\sigma^{q+j+s}$ coincide.

The requirment of localization is automatically satisfied 
in two cases.

\smallskip
\noindent{\sl Case}~1. $k=j+1$. Here always 
$\sigma^{q+j+1}=\sigma^{q+k}$.

\smallskip
\noindent{\sl Case}~2. In a simplicial complex $K$ the 
simplexes $\sigma_1^{q+k}$ and $\sigma_2^{q+k}$ coincide 
if they intersect in a face of dimension greater than 
$q+ j$.

In the examples already considered in this work (above) 
we have always had $q=0$, that is, functions $\psi$ were 
defined on vertices. Localization corresponded to cases~1 or~2.

\example{A trivial example}
Let $k=1,q=0,j=0$. We have the usual Abelian connection, 
sitting on edges of the complex~$K$. In fact, from~(170)
$$
c_{\sigma^1:\sigma_1^0}\psi_{\sigma_1^0}+c_{\sigma^1:\sigma_2^0}
\psi_{\sigma^0_2}=0
$$
we arrive in this case at the operator of multiplication
$$
L_{12}=-\frac{c_{\sigma^1:\sigma^0_1}}{c_{\sigma^1:\sigma^0_2}}\:
\psi_{\sigma_1^0}\longmapsto\psi_{\sigma_2^0}=L_{12}(\psi_{\sigma_1^0}).
$$
\endexample

\example{Example~13}
a) Let $q=0,k=2,j=1$ and let $K$ be a triangulation of a 
two-dimensional surface. Here we have one equation. This 
situation was studied above (see~\S6).

For $K=\mathbb R^2$ with an equilateral triangular lattice 
we paid particular attention to the case of `zero local 
curvature' where a solution of the equation $Q^+\psi=0$ 
reduces to a difference equation of the second order on 
the straight line (see Example~11). According to a hypothesis 
of the authors, this situation arises in some interesting 
cases when the point $\lambda=0$ is supposedly the lower 
bound of the continuous spectrum of the Schr\"odinger 
operator~(123). On surfaces with non-trivial topology the 
case of zero local curvature leads to the global monodromy 
defined on the group~$\pi_1$.

b) Let $q=0,j=1,k=2$ and let $K$ be the `black part' (or 
`white part') of the black and white triangulation of a 
surface. Here we also have one equation $Q^+\psi=0$, but we 
do not have local curvature, since black triangles are adjacent 
to each other at vertices (there are no common edges). 
Therefore there are no non-trivial paths.

c) Let $q=0,j=0,k=2$, and let $K$ be the same as in~b). 
Here we have curvature.
\endexample

\example{Example~14}
Let $q=0$, $k=3$, and let $K$ be the lattice of regular 
tetrahedra in~$\mathbb R^3$. We have $K=K_1\cup K_2$, 
where $K_1$ is the `black' part, $K_2$ is the `white' 
part, and $K_1\cap K_2$ is a one-dimensional skeleton. Adjacency 
of two tetrahedra in $K$ is possible along no more than an 
edge, and adjacency of tetrahedra inside $K_1$ (or~$K_2$) is 
only possible at a vertex. We have already considered the 
following cases~a),~b), see~\S7:

a) $j=2$, complex $K_1$ (or~$K_2$). There is no local 
curvature here.

b) $j=2$, complex~$K$. There is also no local curvature here, 
since there are no paths~$\gamma$. Nevertheless, there is a 
non-trivial condition for consistency (153) which ensures 
an extensive space of solutions of the equation $Q^+\psi=0$. 
This particular analogue of curvature has not been studied in 
a general form.

c) $j=1$, complex~$K$. Here the concept of local curvature 
arises, since we have many paths~$\gamma$. The system (170) 
has a solution in the whole of~$\mathbb Z^3$, which is 
uniquely determined by an arbitrary pair of values of 
$\psi$ at the vertices of any edge if and only if the 
following condition is satisfied: eight equations, 
corresponding to each set of four tetrahedra, a pair of 
black and a pair of white, from the star of one vertex, 
such that both black tetrahedra adjoin each of the white 
ones along an edge, are linearly dependent. This assertion 
means that our system in the star of every vertex has a 
two-dimensional local space of solutions. Geometrically, 
`curvatures' that obstruct this solution correspond on the 
border of the star---a sphere~$S^2$ divided into triangles 
and squares---just to squares. In all we have six conditions 
at each vertex (the number of squares), but one of them is 
dependent.

d) Let $q=0,j=0$, and let $K$ be~$K_1$. There are paths here 
composed of tetrahedra, linked along vertices. The space 
$\mathbb R^M$ for $j=0$ is one-dimensional, and we arrive 
only at Abelian connections. This is true for $j=0$ in any 
simplicial complexes for any~$q$.
\endexample

\proclaim{Lemma~21}
A simplicial connection of type $(q,j,k)$ that satisfies the 
conditions for non-degeneracy and localization defines a 
multiplicative curvature transformation
$$
R_\gamma\:\mathbb R^M\to\mathbb R^M,\qquad
R_{\gamma_1\gamma_2}=R_{\gamma_1}R_{\gamma_2},\qquad
R_{\gamma^{-1}}=R_\gamma^{-1}
\tag 174
$$
for any path~$\gamma$,
$$
\gamma=\sigma_1^{q+k}\sigma_2^{q+k}\dotsb\sigma_N^{q+k},
$$
where $\sigma_N^{q+k}=\sigma_1^{q+k}$ and the intersection 
$\sigma_s^{q+k}\cap\sigma_{s+1}^{q+k}$ is a face of dimension 
greater than or equal to $q+j$ for all $s=1,\dots,N$.
\endproclaim

The proof of the lemma is easily obtained from the previous one.

We now consider real discrete self-adjoint operators
$$
(L\psi)_{\sigma_2^q}=\sum_{\sigma_1^q}b_{\sigma_1^q:\sigma_2^q}\psi_
{\sigma_2^q},
\tag 175
$$
acting on functions of $q$-simplexes of the complex~$K$. 
We consider only `operators of the second order', where 
the coefficients $b_{\sigma_1^q:\sigma_2^q}$ are different 
from zero only for `nearest neighbours'.

\definition{Definition~16}
The simplexes $\sigma_1^q$, $\sigma_2^q$ are $k_+$-nearest 
($k_-$-nearest) if this pair is contained in some simplex 
$\sigma^{q+k_+}$ (respectively, if the intersection 
$\sigma_1^q\cap\sigma_2^q$ is a simplex~$\sigma^{q-k_-}$).
\enddefinition

We can represent the operator $L$ as the sum of operators of 
two types $L=L_++ L_-$. We consider these types 
separately;
$$
(L_+\psi)_{\sigma_2^q}=\sum_{\sigma_1^q,\sigma_2^q\subset
\sigma^{q+k_+}}
b_{\sigma_2^q:\sigma^{q+k_+}:\sigma_1^q}\psi_{\sigma_1^q},
\tag {type \mbox{$+$}}
$$
$$
(L_-\psi)_{\sigma_4^q}=\sum_{\sigma_3^q\cap\sigma_4^q\supset
\sigma^{q-k_-}}b_{\sigma_4^q:\sigma^{q-k_-}:\sigma_3^q}
\psi_{\sigma_3^q},
\tag {type \mbox{$-$}}
$$
where $\sigma_1^q,\sigma_2^q$ are $k_+$-nearest and 
$\sigma_3^q,\sigma_4^q$ are $k_-$-nearest. Let 
$k_+\,{=}\,k_-\,{=}\,k$. We consider an operator $Q^+$ 
of the same type as in the definition of a simplicial connection, 
and the operator~$Q_1$:
$$
(Q^+\psi)_{\sigma^{q+k}}=\sum_{\sigma^q}c_{\sigma^{q+k}:\sigma^q}
\psi_{\sigma^q}\,,
\quad
(Q_1\psi)_{\sigma^{q-k}}=\sum_{\sigma^q}c_{\sigma^q:\sigma^{q-k}}
\psi_{\sigma^q}\,.
$$
Here the images are vector-functions, and the $\psi$ are scalars.

\definition{Definition~17}
1) A {\it factorization of the first type} is a representation 
of $L_+$ in the form
$$
L_+=QQ^++w,
\tag 176
$$
where $Q^+$ is the operator adjoint to $Q$ and $w=w_{\sigma^q}$ 
is the operator of multiplication by a numerical function.

A {\it special factorization of the first type} is a 
representation~(176), where~$w= \const$.

2) A {\it factorization of the second type} is a 
representation of the operator $L_-$ in the form
$$
L_-=Q_1^+Q_1+v,
\tag 177
$$
where $v$ is the operator of multiplication by a function 
of $\sigma^q$ and the operator $Q^+_1$ is adjoint to~$Q_1$.

A {\it special factorization of the second type} is the 
case when $v= \const$.

3) A Laplace transformation of the operators $L_\pm$ is 
defined by the formulae
$$
\aligned
L_+&\longmapsto\widetilde L_+=Q^+w^{-1}Q+1,
\\
L_-&\longmapsto\widetilde L_-=Q_1v^{-1}Q_1^++1,
\endaligned
\tag 178
$$
and their zero eigenvectors are transformed by well-known 
formulae into the eigenvectors of~$\widetilde L_\pm$:
$$
\begin{aligned}
&\psi^+\longmapsto\widetilde\psi^+=Q^+\psi^+,\qquad
\widetilde L_+\widetilde\psi^+=0,\qquad L_+\psi^+=0,
\\
&\psi^-\longmapsto\widetilde\psi^-=Q_1\psi^-,\qquad
\widetilde L_-\widetilde\psi^-=0,\qquad L_-\psi^-=0.
\end{aligned}
$$
Obviously, the operator $\widetilde L_+$ is written 
analogously to $L_-$, where $\widetilde Q^+=v^{-1/2}Q^+$. 
These formulae assume that $w\ge0$,
$v\ge0$, although this does not play a serious role here.
\enddefinition

Detailed studies have been devoted above to the 
investigation of factorizations and Laplace transformations 
in particular cases. Here we consider only the simplest 
example, that is, a standard discrete real Schr\"odinger 
operator.

\example{Example~15}
Let $q=0$, $k=1$, $j=0$. We have an operator defined 
on functions of vertices
$$
L=L_+,\qquad
(L\psi)_{\sigma_2^0}=\sum_{\sigma_1^0\cup\sigma_2^0=
\partial\sigma^1}b_{\sigma_2^0:\sigma^1:\sigma_1^0}
\psi_{\sigma_1^0}+
b_{\sigma_1^0}\delta(\sigma_2^0,\sigma_1^0)
\psi_{\sigma_1^0}.
$$
(The last term is present only for $\sigma_2^0=\sigma_1^0$; 
this is shown by the $\delta$-function.)

We can always factorize such an operator in the form 
(we always suppose that
$b_{\sigma_2^0:\sigma^1:\sigma_1^0}\ne 0$)
$$
L=QQ^++w,
\tag 179
$$
where $w$ is some function. This factorization is not unique.
\endexample

\example{Question}
Is there always a `special factorization', where $w=\const$? 
It exists for the complex $K=\mathbb Z^1$, see~\S3, but in 
the general case the answer is not clear (we assume that 
$b_{\sigma^0}\ge0$).
\endexample

\remark{Remark~$19$}
Earlier (see~\S8) in connection with the Laplace transformation 
on surfaces with black-white triangulation we considered yet 
another form of the factorization: let the complex $K$ and 
two subcomplexes $K_1$,~$K_2$ be given, where $K_1\cap K_2$ 
is a $(q-1)$-dimensional skeleton of the complex~$K$, and the 
operator $L$ acts on functions of $\sigma^q\subset K_1$ 
(white $q$-simplexes):
$$
(L\psi)_{\sigma^q_2}=\sum_{\sigma_1^q}b_{\sigma_1^q:\sigma_2^q}\psi
_{\sigma_1^q},\qquad
\sigma_1^q,\sigma_2^q\subset K_1.
$$
The notion of the `nearest $k$-neighbourhood' is defined for 
simplexes from $K_1$ by means of faces of dimension $q-k$, 
which are common for them with an arbitrary `black' $q$-simplex 
from~$K_2$. The factorization is sought in the form~(179), 
where $w$ is the operator of multiplication by a function, 
and $Q^+$ is an operator of the form
$$
(Q^+\psi)_{\overline\sigma^q}=\sum c_{\overline\sigma^q:\sigma^{q-k}:
\sigma^q}\psi_{\sigma^q},
\tag 180
$$
where $\overline\sigma^q\subset K_2$, $\sigma^q\subset K_1$,
$\sigma^{q-k}\subset\overline\sigma^q\cap\sigma^q$. 
Analogously, the operators $\overline L$ act on `black' 
$q$-simplexes $\overline\sigma^q\subset K_2$ and the 
factorization is sought in the form
$$
\overline L=Q^+Q+v.
$$
Previous formulae define Laplace transformations and also 
special Laplace transformations if the potentials $v,w$ are 
constant.
\endremark

\example{Example~16}
Let $K$ be a $q$-dimensional triangulated manifold with 
black-white coloured $q$-dimensional simplexes, $K=K_1\cup K_2$, 
and let $K_1\cap K_2$ be a \linebreak$(q-1)$-dimensional 
skeleton. For $k=1$ we have $\sigma^{q-1}$ as a common face 
of exactly two simplexes $\overline\sigma^q,\sigma^q$, that 
is, black and white. By analogy with the case $q=2$ (see~\S8), 
the condition for factorization depends on the 
$N(\sigma^{q-2})$-multiplicity of $(q-2)$-dimensional 
simplexes $\sigma^{q-2}\subset K$, which is equal to 
the number of adjacent $(q-1)$-dimensional 
$\sigma^{q-1}\supset\sigma^{q-2}$. This number is even, 
$N\ge4$. For $\sigma_j^q\subset K_1$, 
$\overline\sigma_j^q\subset K_2$ we have:
\roster
\Item"a)" $b_{\sigma_1^q:\sigma_2^q}
=c_{\sigma_1^q:\sigma_1^{q-1}:\overline\sigma^q}
c_{\sigma_2^q:\sigma_2^{q-1}:\overline\sigma^q}$, \
$\sigma^{q-2}=\sigma_1^{q-1}\cap\sigma_2^{q-1}$, \
$N(\sigma^{q-2})\ge6$;
\Item"b)" $b_{\sigma_1^q:\sigma_2^q}
=c_{\sigma_1^q:\sigma_1^{q-1}:\overline\sigma_1^q}
c_{\sigma_2^q:\sigma_2^{q-1}:\overline\sigma_1^q}
+c_{\sigma_1^q:\sigma_3^{q-1}:\overline\sigma_2^q}
c_{\sigma_2^q:\sigma_4^{q-1}:\overline\sigma_2^q}$, \
$\sigma^{q-2}=\sigma_1^{q-1}\cap\sigma_2^{q-1}\cap
\sigma_3^{q-1}\cap\sigma_4^{q-1}$, \
$N(\sigma^{q-2})=4$.
\endroster

Let us consider a): the number of equations is equal to the 
number of \linebreak$(q-2)$-dimensional faces of the simplex 
$\overline\sigma^q$, that is, $l=q(q+1)/2$. The number of 
unknowns is $m(q+1)$, where $m$ is the dimension of the 
vector $c_{\sigma^a:\sigma^{q-1}:\overline\sigma^q}$. Thus 
we have the condition $m(q+1)\ge q(q+1)/2$ or $m\ge q/2$. 
The scalar factorization ($m=1$) needs to satisfy additional 
conditions for~$q\ge3$.
\endexample

\example{Example~17}
We again consider $K$, a $(q+2)$-dimensional manifold 
with `black-white' coloured $(q+2)$-simplexes, $K=K_1\cup K_2$, 
and $K_1\cap K_2$, a $(q+1)$-skeleton.\linebreak
Suppose we are given an operator of 
some other type, acting on functions of \linebreak$q$-simplexes 
from~$K$
$$
(L\psi)_{\sigma_2^q}=\sum_{\sigma_1^q}b_{\sigma_2^q:\sigma_1^q}
\psi_{\sigma_1^q}\,,
$$
and we seek a `white' factorization~(179), where $w$ is 
the operator of multiplication by a function, and
$$
(Q^+\psi)_{\sigma^{q+2}}=\sum_{\sigma^q}c_{\sigma^{q+2}:\sigma^q}
\psi_{\sigma^q}\,,\qquad
\sigma^{q+2}\subset K.
$$
We arrive at the relation
$$
b_{\sigma_1^q:\sigma^q_2}=c_{\sigma^{q+2}:\sigma_1^q}
c_{\sigma^{q+2}:\sigma_2^q}\,.
$$
The number of unknowns is equal to $m(q+2)(q+3)/2=lm$ (if 
$c_{\sigma^{q+2}:\sigma_1^q}$ is an $m$-vector) and the 
number of equations is equal to $l(l-1)/2$. The factorization 
condition thus has the form
$$
lm\ge l(l-1)/2,
\tag 181
$$
$(q+2)(q+3)/2=l$, $m\ge(l-1)/2$.

For $q=1,k=2$ we see that $m=3$ or $5$ is acceptable. In these 
cases factorization is always possible in the form~(179). The 
possibility of the special factorization $w=\const$ is of interest: 
one would like to clarify this question. If the factorization is 
special, then the question of zero modes of the operator $L=QQ^+$ 
is of interest, that is, the question of the solubility in the 
space $\psi\in\script L_2$ of the equation
$
Q^+\psi=0
$.
For $m=3$ and $m=5$ equations of this type determine the 
connections (if the conditions for non-degeneracy and localization 
are satisfied). Paths along which `parallel transport' is realized 
and curvature is defined consist of 3-simplexes ($q=1$) 
adjoining each other along edges ($m=5$) or along faces ($m=3$). 
The requirement of localization is automatically satisfied for the 
case $m=5$, $K=K_1$ (white \linebreak tetrahedra).

If $m=3$, then purely white paths do not exist. The curvature 
is defined by the pair of equations: $Q_1^+\psi=0$ (white part), 
$Q_2^+\psi=0$ (black part), that is, $K=K_1\cup K_2$.
\endexample

Now suppose we are given an arbitrary simplicial connection on 
a complex $K$ of type~$(q,j,k)$.

\definition{Definition~18}
The {\it local curvature of the vertex} $\sigma^0\subset K$ is 
the curvature of this connection on special paths of the 
subcomplex~$K_P$, which is the simplicial star of the 
vertex~$\sigma^0$: we take all simplexes $\sigma_c^q$ with 
vertex $\sigma^0$, $\sigma^q\subset K_P$, paths of the form
$$
\gamma=(\sigma_1^{q+k},\sigma_2^{q+k},\dots,\sigma_N^{q+k}=
\sigma_1^{q+k}),
\tag 182
$$
such that $\sigma_j^{q+k}=\sigma^0\sigma_j^{q-1+k}$.

A transformation $R_\gamma$ determined only by paths $\gamma\subset 
K_P$ (182) is called the {\it local curvature of the simplicial 
connection at the point} $\sigma^0\subset K$.
\enddefinition

\remark{Hypothesis}
For manifolds $K$ the curvature is trivial if the local curvatures 
of all the vertices are trivial (we note that in a number of cases 
above $K$ was not a manifold).
\endremark

\end{document}